\documentclass[hidelinks,a4paper,10pt,onecolumn, switch]{article}
\pdfoutput=1

\usepackage{fancyhdr}
\fancypagestyle{plain}{%
	\fancyhf{}
	\fancyfoot[R]{\thepage}
	\fancyfoot[L]{\small \textit{Preprint}\\ \texttt{arXiv}}
}

\usepackage[affil-it]{authblk}
\makeatletter
\def\@maketitle{%
	\newpage
	\null
	\vskip 2em%
	\begin{center}%
		\let \footnote \thanks
		{\Large\bfseries \@title \par}%
		\vskip 1.5em%
		{\normalsize
			\lineskip .5em%
			\begin{tabular}[t]{c}%
				\@author
			\end{tabular}\par}%
		\vskip 1em%
		{\normalsize \@date}%
	\end{center}%
	\par
	\vskip 1.5em}
\makeatother

\usepackage[a4paper, total={17.5cm,24cm}]{geometry}

\usepackage[font=normal,labelfont=bf]{caption}

\usepackage{sectsty}

\usepackage{titlesec}
\titlespacing\section{0pt}{12pt plus 3pt minus 3pt}{1pt plus 1pt minus 1pt}
\titlespacing\subsection{0pt}{10pt plus 3pt minus 3pt}{1pt plus 1pt minus 1pt}
\titlespacing\subsubsection{0pt}{8pt plus 3pt minus 3pt}{1pt plus 1pt minus 1pt}

\titleformat{\section}{\normalfont\large\bfseries}{\thesection}{1em}{}

\titleformat{\subsection}{\normalfont\normalsize\bfseries}{\thesubsection}{1em}{}

\titleformat{\subsubsection}{\normalfont\normalsize}{\thesubsubsection}{1em}{}

\titleformat{\paragraph}[runin]{\normalfont\normalsize\itshape}{\theparagraph}{1em}{}

\usepackage{mathtools,upgreek,amsmath,amssymb}

\usepackage[utf8]{inputenc}	
\usepackage[T1]{fontenc}	
\usepackage{xcolor}		
\usepackage[linktoc=page,
colorlinks=true,	
linkcolor=blue,
citecolor=blue,
urlcolor=blue]{hyperref}     
\usepackage{booktabs} 		
\usepackage{nicefrac}		
\usepackage{microtype}		
\usepackage{lineno}		
\usepackage{float}			
\usepackage{svg}

\usepackage{siunitx}[=v2]
\usepackage{nicefrac,multirow}
\usepackage{import}
\usepackage{mathrsfs}

\usepackage{subcaption}
\usepackage{algorithm}
\usepackage{algpseudocode}
\usepackage{enumitem} 
\setlist[enumerate]{label*=\arabic*.}

\DeclareMathOperator{\tr}{tr}

\DeclareMathOperator{\cl}{cl}

\newcommand{\inte}[3]{\int \limits_{ #1} #2 \; \mathrm{d} #3}

\newcommand{\nabr}{\nabla_{\ve{X}}}

\newcommand{\diffp}[2]{\frac{\partial #1}{\partial #2}}

\newcommand{\ve}[1]{\boldsymbol{#1}} 
\newcommand{\te}[1]{\mathbf #1}

\newcommand{\nv}{\mathrm}
\newcommand{\comma}{\hspace{3mm} \text{,}}

\newcommand{\point}{\hspace{3mm} \text{.}}

\newcommand{\psii}{\overline \psi}

\newcommand{\psiinno}{\psii^\nv{PANN}_0}
\newcommand{\psiinon}{\psii^\nv{NN}}
\newcommand{\piso}{\overline P}

\newcommand{\fiso}{\overline{\te F}}
\newcommand{\ciso}{\overline{\te C}}

\newcommand{\bref}{\mathcal{B}_{0}}

\newcommand{\omref}{\varOmega_{0}}
\newcommand{\omrefe}{\varOmega_{0}^{e}}
\newcommand{\domref}{\partial\varOmega_{0}}
\newcommand{\domreft}{\partial\varOmega_{0,\hat{\ve T}}}

\newcommand{\rs}{\mathbb{R}}
\newcommand{\rsnn}{\rs_{\geq 0}}
\newcommand{\rsp}{\rs_{> 0}}
\newcommand{\so}{\mathbb{SO}(3)}

\newcommand{\tes}{\mathcal{T}^2}
\newcommand{\tef}{\mathcal{T}^1}
\newcommand{\tesp}{\mathcal{T}^2_+}
\newcommand{\tesu}{\mathcal{T}_\nv{u}^2}

\newcommand{\pp}{\tilde p}

\usepackage{multicol}

\DeclareMathOperator{\diag}{diag}
\DeclareMathOperator{\cof}{cof}

\newcommand{\ici}{\overline{I}_1}
\newcommand{\iici}{\overline{I}_2}

\newcommand{\Ldom}{ \mathcal{L}_{\! \mathscr{F}}}
\newcommand{\Lbound}{ \mathcal{L}_{\! \mathscr{L}}}

\definecolor{pltred}{RGB}{214,39,40}
\definecolor{darkgray176}{RGB}{176,176,176}
\definecolor{pltorange}{RGB}{255,127,14}
\definecolor{darkslategray51}{RGB}{51,51,51}
\definecolor{pltgreen}{RGB}{44,160,44}
\definecolor{lightgray204}{RGB}{204,204,204}
\definecolor{pltblue}{RGB}{31,119,180}

\newcommand{\smallbullet}{\raisebox{0.1em}{\scalebox{0.6}{$\bullet$}}}

\usepackage{layouts}
\DeclareEmphSequence{\bfseries}

\usepackage{mdframed}

\usepackage{amsthm}

\theoremstyle{definition}

\theoremstyle{remark}

\title{
Reliable training of neural hyperelastic models via full-field data}
\begin{document}

	\author[a]{Konrad Friedrichs}
	\author[a]{Franz Dammaß}
	\author[a]{Karl A. Kalina}
	\author[,a,b]{Markus Kästner\thanks{Contact: \texttt{markus.kaestner@tu-dresden.de} }}
	\affil[a]{Institute of Solid Mechanics, TU Dresden, Germany}
	\affil[b]{DCMS -- Dresden Center for Computational Materials Science, Dresden, Germany}
	\date{}
	\maketitle
	
	\begin{abstract}
    We present a systematic investigation of the robustness and limitations of equilibrium gap-based calibrations for hyperelastic physics-augmented neural networks (PANNs), where we consider the special case of isotropic and polyconvex PANNs.
    In full-field parameterizations, it is commonly assumed that the displacement field is captured with sufficient spatial resolution for an accurate evaluation of the deformation field, and that the specimen is thin enough for plane stress to hold to a good approximation.
    Since these assumptions are never ideally satisfied in real experiments, we investigate, using synthetically generated data, how severely an under-resolved surface measurement and a non-negligible specimen thickness can affect the model parameterization.
    Furthermore, we perform calibration on real experimental data for a set of inhomogeneous specimen geometries.
    We show that the coverage of the admissible deformation states during calibration governs the ability of a model to generalize to unseen geometries and load cases; this ability can be improved further by appropriate combinations of specimens.
    Accurately depicting the material behavior underlying this rich data, however, requires a sufficiently flexible constitutive model, for which PANNs are well suited.
    Yet a rich coverage of deformation states alone is not sufficient: unless the calibration data comprise biaxial-tension-like states, models that include the second deformation invariant extrapolate unphysically towards equi-biaxial tension, whereas restricting the PANN to the first invariant remains reliable.
    \\
    \vspace{5mm}
	\noindent\\
	\textbf{Keywords: }\textendash~Incompressibility~\textendash~Full-field calibration~\textendash~Equilibrium gap method~\textendash~Finite deformations~\textendash~Plane stress assumption~\textendash~Physics-augmented neural networks
	
	\end{abstract}
	
\section{Introduction}
\label{sec:intro}
The mechanical characterization of materials undergoing finite deformations remains an important challenge in computational mechanics.
Rubber-like materials typically exhibit highly nonlinear and nearly incompressible elastic behavior that demands both flexible constitutive models and reliable calibration strategies.
On the modeling side, machine learning approaches have recently provided the versatility to capture such complex material behavior, which could often not be achieved with classical model formulations. 
On the calibration side, constitutive model parameters have been traditionally identified from standard experiments, such as uniaxial tension, biaxial tension, and pure shear, where homogeneous deformation and stress can be assumed throughout the specimen, cf.~\cite{treloar1944, jones1975, kawabata1981}.
While elegant in their simplicity, such approaches require multiple complementary experiments to adequately capture the material response under multiaxial loading, and are limited to narrow samplings of deformation states \cite{steinmann2012, marckmann2006, ricker2023, dammass2025b}.
The advent of full-field measurement techniques, combined with digital image correlation (DIC), has transformed the field of experimental mechanics by providing spatially resolved displacement fields across entire specimen surfaces \cite{chevalier2001, chrysochoos2010, belloni2019}.
In contrast to conventional experiments, inhomogeneous specimen geometries can be exploited to cover a wide range of deformation and stress states, thereby shifting the field of constitutive modeling from a limited-data to a large-data regime.
While inhomogeneous in-plane displacement fields are attainable at points across the domain using, e.g., camera systems in combination with DIC \cite{pierron2021}, the associated stress fields are generally not measurable.
As a remedy, the identification of constitutive model parameters can be carried out by reproducing the experiment numerically and solving an inverse problem.\par
In what follows, we give an overview of the most common full-field calibration strategies, with particular focus on data-driven constitutive models.

\subsection{Literature overview}
In their pioneering work, \cite{ghaboussi1991} were the first to use neural networks (NNs) for describing constitutive behavior in the early 1990s.
The most recent advancements in computational power and machine learning have revived the pursuit of this field, and thereby paved the way for the current rise of data-driven techniques in mechanics, cf.~\cite{bock2019,dornheim2024,fuhg2025} for comprehensive reviews.
A major advancement in neural constitutive modeling is the integration of essential physical concepts, an idea that has received various designations in the literature \cite{raissi2019,asad2022, klein2022,linden2023,aldakheel2025,kalina2023,masi2021,linka2021,geuken2025a}.
For this purpose, physical considerations may be incorporated either strongly, that is, through network architectures tailored to the problem, or in a weak sense, by formulating problem-specific loss functions for training, which both substantially contribute to the extrapolation capabilities~\cite{fuhg2023, linden2023, masi2021}. 
For elasticity, an established approach is to construct neural network architectures to define invariant-based hyperelastic potentials \cite{bahmani2024, benady2024a, fuhg2022, klein2022, linden2023, linka2021, peirlinck2024, tac2024}.
In the special case of isotropic, incompressible hyperelasticity, as is the scope of this work, the two invariants $\ici$ and $\iici$ of the isochoric right Cauchy-Green deformation tensor constitute the corresponding invariant basis~\cite{dammass2025a,dammass2025b}.
Therein, polyconvex architectures are a popular further ingredient \cite{bahmani2024, chen2022, dammass2025a,dammass2025b, geuken2025a, jadoon2025a, klein2022, tac2024, vijayakumaran2025}, often achieved through input-convex neural networks~\cite{amos}, thereby guaranteeing rank-one convexity and thus ellipticity \cite{schroder2010}, and improving the extrapolation behavior \cite{kalina2024, linden2023}. 
The increased flexibility of this modeling paradigm enables a shift away from limiting, classical formulations, thereby being able to handle the large-data regime made available by full-field measurements.
\par
Calibrating such data-driven models from full-field data requires a suitable inverse identification strategy.
Over the past decades, several distinct methodological frameworks for the inverse identification of constitutive parameters have been established.
A comprehensive overview of full-field-based identification techniques, together with a systematic comparison of the most prominent approaches, is provided in \cite{avril2008}.
These comprise the finite element model updating (FEMU) method \cite{avril2008,romer2025}, the constitutive equation gap method (CEGM), the virtual
fields method (VFM) \cite{grediac1989,pierron2012}, the equilibrium gap method (EGM) \cite{claire2004}%
\footnote{\label{foot:EGM_VFM}The EGM may be regarded as a special case of the VFM, in which the virtual fields are generated from locally supported finite element test functions \cite{avril2008,romer2025}.} and the reciprocity gap method (RGM).
These frameworks differ primarily in whether a solution of the forward problem is required.
Methods that require such a solution, such as FEMU, do not strictly rely on full-field displacement data from experiments, although incorporating such data, where available, can increase the identification accuracy.
Approaches such as the EGM and the VFM, in contrast, operate directly on full-field
displacement data that needs to be available across, at least to a good approximation, the entire domain, from which deformation measures and stresses are determined without solving the underlying boundary value problem.
This renders them considerably more computationally efficient, yet also more susceptible
to measurement noise. 

Over the past few years, inverse parameter identification methods have been transferred towards data-driven applications, and novel data-driven strategies have emerged in parallel. 
A prominent example which extends the EGM/VFM methodology is the EUCLID (Efficient Unsupervised Constitutive Law Identification
\& Discovery) framework.
Within this scheme, both the material parameters and appropriate constitutive models, which are selected from a predefined model catalog, are simultaneously identified in an unsupervised learning setting via sparse regression \cite{flaschel2021,flaschel2023}. 
Experimental validation of the approach using 2D DIC data has been demonstrated in \cite{abbasi2026a}.
Several extensions of this framework have since been proposed. 
In NN-EUCLID \cite{thakolkaran2022}, the discrete model catalog is substituted by a neural network-based constitutive model, whose weights and biases are learned directly from data. 
Building on this idea, CANN-EUCLID \cite{alheit2026} performs calibration of a constitutive artificial neural network~(CANN) following the formulation of \cite{linka2023}. 
A comparable strategy is adopted in \cite{moon2026} for the discovery of neural network-based yield surfaces. 
Furthermore, the same underlying methodology has been leveraged in \cite{thakolkaran2025} to train Kolmogorov-Arnold networks. 
In \cite{meng2025}, the approach is applied to 3D displacement fields acquired during a bulge inflation test, while \cite{bourdyot2026} demonstrates its use with 3D digital volume correlation data obtained for a printed structure. 
Building on this direction, \cite{ferreira2026} employs the EGM to calibrate a range of elasto-plastic constitutive models, including formulations based on recurrent neural networks (RNNs).
The integration of RNN-based constitutive models into the VFM was previously investigated in \cite{lourenco2024}.
Furthermore, a variant specifically tailored to anisotropic elasticity is proposed in \cite{li2026}, and dedicated frameworks addressing heterogeneous samples have been developed in \cite{shi2025a,tac2026,chaurasiya2026}.
The NN-mCRE (neural network modified Constitutive Relation Error) approach extends the classical mCRE framework by replacing the conventional constitutive model with a PANN \cite{benady2024,benady2024a}. 
Several complementary strategies have emerged along similar lines.
In \cite{wu2025,tan2026}, PANNs are combined with FEMU, while \cite{wiesheier2024,wiesheier2026} formulate and calibrate spline-based constitutive models within the same FEMU setting. 
In a comparable spirit, \cite{knipper2026} employs FEMU to train CANNs for isotropic, incompressible hyperelasticity. 
This methodology has been further extended to inelastic regimes: NN-based elasto-plastic models are calibrated via FEMU in \cite{gavris2025}, and neural cohesive zone models in \cite{gavris2026}.
Most recently, the EGM has also been employed for the calibration of viscoelastic PANN models in~\cite{riemer2026a}.

\subsection{Objectives and contributions of this work}
Despite these advances, the systematic calibration of physics-augmented neural network constitutive models from full-field experimental data via the EGM at finite strains has received comparatively little attention so far, and how reliably such models extrapolate beyond the calibrated regime remains less well understood.

Further, a key assumption in many of these works is that the thin sheet specimens employed in the experiments experience in-plane stress states, which allows the out-of-plane deformation to be inferred from the measurements on the surface, either via the incompressibility constraint, or through fulfillment of the stress boundary condition.
While this assumption is widely adopted, its validity, particularly near specimen boundaries, geometric features, or for non-negligible specimen thickness, where three-dimensional stress states may arise, has received limited systematic investigation.
Additionally, many studies rely on synthetic full-field data, in which the relevant field quantities are available at a virtually arbitrarily fine spatial resolution.
However, it remains uncertain how strongly these intricate methodologies suffer from limited resolutions of the surface displacements, as is the case in real experimental settings, and how this propagates to the model calibration.

Building on these developments, we employ physics-augmented neural networks as the constitutive modeling paradigm throughout the present study, so that the flexibility required
to depict the material behavior at all deformation states covered by the
full-field data is available by construction.
Within the context of incompressible hyperelasticity, the present contribution therefore addresses the following questions concerning the equilibrium gap method:
\begin{itemize}[label=\smallbullet, itemsep=3pt, topsep=4pt, parsep=0pt]
    \item How does the thickness of a specimen affect the validity of the plane stress assumption?
    \item Which surface resolution of the displacement field is sufficient for a robust calibration?
    \item How can different specimen geometries be leveraged to achieve a broad representation of deformation states during calibration?
\end{itemize}

For this purpose, the remainder of this paper is organized as follows.
Section 2 presents the theoretical foundations, including the PANN model and the EGM formulation for incompressible hyperelasticity. 
In Section 3, we investigate the necessity of a high surface resolution and the limits of the plane stress assumption employing synthetic data.
Section 4 then turns to the calibration via experimental data, examining the extrapolation capabilities of the identified PANN models together with the roles of specimen geometry and model complexity.

\noindent\textbf{\textit{Notation.}} Within this paper, italic symbols are used for scalar quantities ($J, \psi$) and bold italic symbols for first-order tensors ($\boldsymbol{u} \in \tef$). 
For second-order tensors, bold upright letters ($\mathbf{P}, \boldsymbol{\upsigma} \in \mathcal{T}^2$) are used.
We consider a Cartesian coordinate frame in the Euclidean space $\mathbb R^3$,
with its standard basis~$(\ve e_1, \ve e_2, \ve e_3)$, such that tensors can be identified with the corresponding components.
The symbol $\otimes$ denotes the dyadic product, and the Einstein summation convention applies.
Transpose and inverse of $\mathbf{t} \in \mathcal{T}^2$ are given by $\mathbf{t}^\top$ and $\mathbf{t}^{-1}$, respectively. Additionally, $\mathrm{tr}\,\mathbf{t}, \, \det \mathbf{t}, \, \mathrm{cof}\,\mathbf{t} = \det(\mathbf{t})\, \mathbf{t}^{-\top}$ are used to indicate trace, determinant and cofactor, respectively.
A second-order tensor $\mathbf{d}$ that is represented by a diagonal matrix with $a_1, a_2, a_3$ on its main diagonal is written as $\mathbf{d} = \mathrm{diag}(a_1, a_2, a_3)$. 
Furthermore, relevant sets of tensors are denoted as 
$
\mathcal{T}^2_\text{s} = \left\{ \mathbf{t} \in \mathcal{T}^2 \,\middle|\, \mathbf{t} = \mathbf{t}^\top \right\}, \quad
\mathcal{T}^2_+ = \left\{ \mathbf{t} \in \mathcal{T}^2 \,\middle|\, \det \mathbf{t} > 0 \right\}, \quad
\mathcal{T}^2_\text{u} = \left\{ \mathbf{t} \in \mathcal{T}^2 \,\middle|\, \det \mathbf{t} = 1 \right\} \ \mathrm{and}\quad
\mathcal{T}^2_\text{s>} = \left\{ \mathbf{t} \in \mathcal{T}^2_\text{s} \,\middle|\, \ve v \cdot \te t \cdot \ve v > 0 \ \forall \ve v \in \tef \setminus \! \boldsymbol{\mathit 0} \right\}.
$
Moreover, the special orthogonal group is given by $(\mathbb{SO}(3), \cdot)$ with 
$
\mathbb{SO}(3) = \left\{ \mathbf{Q} \in \mathcal{T}^2 \,\middle|\, \mathbf{Q}^{-1} = \mathbf{Q}^\top \wedge \det \mathbf{Q} = 1 \right\},
$
and the single contraction $\cdot : (\mathbf{C}, \mathbf{D}) \mapsto \mathbf{C} \cdot \mathbf{D} = C_{kl} D_{li} \, \mathbf{e}_k \otimes \mathbf{e}_i$. 
The second-order identity tensor is denoted as $\mathbf{I}$.
The closure of a set is denoted by $\cl\varOmega = \varOmega \cup \partial\varOmega$, where the boundary may be segmented, with the individual parts identified by a superscript, i.e., $\partial\varOmega^\varXi$.
The symbol $\nabr \ve \bullet = \diffp{\bullet}{X_J}   \ve e_J$ 
denotes the nabla operator, with $\diffp{\bullet}{X_J}$ indicating the (weak) partial derivative with respect to $X_J$.
Quantities evaluated at a given time increment are marked by a left superscript $n \in \{1, \dots, n_\mathrm{T}\}$, where $n_\mathrm{T}$ denotes the number of increments. 
Quantities associated with a finite element carry a right superscript $e \in \{1, \dots, E\}$, and those associated with a global node a right subscript $\alpha \in \{1, \dots, K\}$, with $E$ and $K$ denoting the number of elements and nodes, respectively.
Finally,
quantities related to the isochoric portion of deformation are marked by an overbar, e.g., $\overline{\psi}$.
For reasons of readability, the arguments of functions are usually omitted within this work, and, in general, no explicit distinction is made between tensor fields and tensors.
    
\section{Fundamentals}
\label{sec:fundamentals}

\subsection{Preliminaries on continuum mechanics}

Let a material body be defined by the bounded, sufficiently smooth domain 
$\mathcal{B}_0 \subset \rs^3$
that it occupies in the undeformed reference state.
For any time $t$ in the time interval of interest $T \subset \rs$, the family of motions
$\ve \chi_t : \cl \mathcal{B}_0 \rightarrow \cl \mathcal{B}_t$, 
$\ve X \mapsto \ve \chi_t(\ve X) = \ve x(\ve X)$, parametric in time, describes a sufficiently smooth and orientation-preserving mapping of material coordinates 
$\ve X \in \cl \mathcal{B}_0 $  
of the reference configuration to corresponding coordinates 
$\ve x \in \cl \mathcal{B}_t$ 
in the current configuration at time~$t$.
To facilitate the formulation of the time-dependent problem, we further introduce a space- and time-dependent representation of the motion mapping 
$\ve \chi : (\ve X, t) \mapsto \ve \chi(\ve X, t) = \ve x(\ve X, t)$,
for which sufficient smoothness in both space and time is assumed.
The deformation gradient $\te F$ and its determinant $J$ are then defined as
$\te F = \nabr \ve \chi$
and
$J = \det \te F$,
where $\nabr \ve \chi$ denotes the gradient of $\ve \chi$ with respect to the reference coordinate $\ve X$.%
\footnote{In accordance with \cite{bertram1989}, we assume that all spatial derivatives, which are formally only defined on the domain $\mathcal{B}_0$, can be smoothly continued to the boundary $\partial \mathcal{B}_0$.}
Following Flory~\cite{flory1961}, at every point, $\te F$ can be decomposed into a volumetric and an isochoric part
$\te F = \te F^\nv{vol} \cdot \fiso$
with
$\fiso = J^{-1/3}\, \te F \in \tesu$, i.e., $\det \fiso = 1$ and
$\te F^\nv{vol} = J^{1/3} \, \te I$
, where $\te I \in \tes$ denotes the identity. 
As a suitable measure of deformation, we introduce the right Cauchy-Green deformation tensor
$\te C = \te F^\top \cdot \te F \in \mathcal{T}^2_\text{s>}$, with its isochoric counterpart
$\ciso = \fiso^\top \! \cdot \fiso = J^{-2/3} \, \te C$.\par
Within the formalism of incompressible hyperelasticity, stress is derived from a smooth Helmholtz free energy function
\begin{equation}
    \psi:\tesp\times\rs \rightarrow \rs \comma \quad (\te F, \pp)\mapsto \psi(\te F, \pp) = \psii\bigl(\fiso(\te F)\bigr)+\pp\bigl(J(\te F)-1\bigr) \ ,
\end{equation}
where $\pp$ denotes a pressure-type variable, and the energy associated with isochoric deformation 
\begin{equation}
    \psii:\tesu \rightarrow\rsnn \comma \quad \fiso \mapsto\psii(\fiso) \ ,
\end{equation}
remains to be specified. 
Accordingly, the first Piola-Kirchhoff stress is defined as $\te P = \partial\psi / \partial \te F = \overline{\te P} + \pp J \te F^{- \top}$, where $\overline{\te P } = \partial\psii/\partial \te F$ denotes the stress associated with isochoric deformation.
For a more detailed introduction to continuum mechanics, the reader is referred to the textbooks \cite{ciarlet1988a,haupt2002,marsden1984,silhavy1997}.

\subsection{Physics-augmented neural network model for incompressible hyperelasticity}
Classical constitutive models for hyperelasticity often lack the flexibility to accurately capture the highly nonlinear responses observed in soft materials such as elastomers or biological tissues. 
Machine learning-based approaches, and neural networks in particular, offer a remedy due to their adaptability to various constitutive behaviors.
In what follows, we adopt the PANN framework for incompressible, isotropic hyperelasticity proposed in \cite{dammass2025a,dammass2025b}. 
The model represents the free energy associated with isochoric deformation $\psii{}^\mathrm{PANN}$ by means of a component-wise non-decreasing fully input-convex neural network $\psii{}^\mathrm{NN}$, cf.~\cite{amos}, with non-negative weights and monotonically increasing activation functions, depending on the polyconvex invariants 
$\ici = \tr \ciso$ and $\iici^{3/2} = (\tr \cof \ciso)^{3/2}$, with an additional correction term $\psiinno \in \rs$ to ensure zero energy in the absence of deformation.
More precisely, we define
\begin{equation}
	\psii{}^\mathrm{PANN} \left(\ici(\te F),\iici^{3/2}(\te F) \right) = \psii{}^\mathrm{NN} \left(\ici(\te F),\iici^{3/2}(\te F) \right) + \psiinno \comma
\end{equation}
and
\begin{equation}
	\psiinno = - \psiinon\left(\ici(\te I),\iici^{3/2}(\te I) \right)  \ .
\end{equation}
In doing so, the following physical principles are satisfied a priori by construction:
\begin{itemize}[label=\smallbullet, itemsep=3pt, topsep=4pt, parsep=0pt]
    \item \textit{Thermodynamic consistency} $\te \piso = \partial{\psii{}^\mathrm{PANN}} \bigl(\te \fiso(\te F)\bigr)/ \partial{\te F}$
    \item \textit{Objectivity} $\psii{}^\mathrm{PANN}(\te Q \cdot \te \fiso) = \psii{}^\mathrm{PANN}(\te \fiso) \ \forall \, \te Q \in \so \, , \  \fiso \in \tesu$ 
    \item \textit{Isotropy} $\psii{}^\mathrm{PANN}( \te \fiso\cdot \te Q )=\psii{}^\mathrm{PANN}(\te \fiso)\ \forall \, \te Q \in \so \, , \  \fiso \in \tesu$
    \item \textit{Non-negativity of the energy} $\psii{}^\mathrm{PANN} (\te \fiso) \geq 0 \ \forall \, \te \fiso \in \tesu$
    \item \textit{Zero energy in the undeformed state} $\psii{}^\mathrm{PANN} (\te I) = 0$
    \item \textit{Stress-free undeformed state} $\te \piso ( \te I) = \te 0$
    \item \textit{Polyconvexity} \cite{ball1976} 
\end{itemize}
By embedding these conditions directly into the network architecture rather than enforcing them weakly through penalty terms during training, the PANN guarantees physically admissible predictions even for unseen deformation states, while retaining the strong expressivity needed to depict complex material behavior.

In the following, we also investigate the consequences of dropping one of the invariants, $\ici$ or $\iici$.
Although the remaining set then no longer constitutes a functional basis, cf.~\cite{riemer2026}, this restriction may improve the generalization behavior of the model, see~\cite{dammass2025b}.
Moreover, we consider different specific network architectures for $\psii{}^\mathrm{NN}$.
We employ the naming convention $\mathrm{PANN}^\mathcal{N}_{I}$ to concisely indicate the network architecture: $\mathcal{N} = (n_1, \dots, n_l) \in \mathbb{N}^l$ collects the number of neurons per hidden layer, where $l \in \mathbb{N}_{>0}$ denotes the number of hidden layers, and $I$ is a placeholder for the considered input invariants, i.e., either $\ici$ or $\iici$, or both.

\subsection{The equilibrium gap method for incompressible hyperelasticity}
\label{sec:VFM}

To formulate the EGM in the first place, we rely on the following common assumptions regarding the experimental setup and available data:
\begin{itemize}[label=\smallbullet, itemsep=3pt, topsep=4pt, parsep=0pt]
	\item The specimen is thin compared to its in-plane characteristic length, thereby rendering plane stress a valid approximation.
    \item For every time step, the global reaction forces are measured at the corresponding boundaries.
	\item A triangulation of the specimen's surface and the corresponding nodes are available in the reference configuration.
	\item In-plane displacements 
    at the nodes are known at every increment.
    \item Based on the triangulation and the nodal displacement values, the field quantities are interpolated using standard finite element approximation.
\end{itemize}
The central idea of the EGM is to enforce the weak formulation of the balance of linear momentum by incorporating terms that penalize deviations from mechanical equilibrium into the loss function for identification of model parameters. 
In other words, equilibrium is enforced at the discrete level for the assembled nodal forces at every increment $\prescript{n}{}{t} \in \{ \prescript{1}{}{t}, \dots, \prescript{n_\mathrm{T} \!}{}{t} \} \approx T, \ n,n_\mathrm{T} \in \mathbb{N} $.
Employing standard finite element interpolation, the reference domain is approximated by
$\cl \bref \approx \cl \omref = \bigcup_{e=1}^{E} \cl \omrefe $, i.e., the union of element sets $\cl \omrefe$.
Likewise, the in-plane displacement and deformation gradient fields are approximated.
To this end, suitable globally defined, locally supported, continuous and piecewise-smooth shape functions $N^\alpha:\cl \omref \rightarrow \rs, \ \ve X\mapsto N^\alpha(\ve X)$ are considered, which are here defined by piecewise-linear Lagrange interpolation polynomials arising from discretization using three-node triangular elements, extruded in the thickness direction by the constant initial specimen thickness~$h_0 \in \rsp$.
Since only the in-plane components of the displacement are available, we let the indices
$a,A,\dots, d,D$ be in $\{1,2\}$, denote the nodal displacements identified from DIC by $\prescript{n }{}{u_a^\alpha}$, and define
\begin{equation}
    \label{eq:disp_F_ansatz}
    \prescript{n }{}{u}_a =   \sum_{\alpha = 1}^K N^\alpha \prescript{n }{}{u}_a^\alpha 
    \qquad \text{and} \qquad
    \prescript{n \!}{}{F}_{aB} = \delta_{aB} +  \sum_{\alpha = 1}^K  \diffp{N^\alpha}{X_B} \prescript{n }{}{u}_a^\alpha \quad .
\end{equation}
Conversely, the out-of-plane component $\prescript{n\!}{}{F}_{33}$ follows directly from $\prescript{n\!}{}{J}=1$ via $\prescript{n\!}{}{F}_{11}, \dots, \prescript{n\!}{}{F}_{22}$.
Likewise, since plane stress is assumed, $\prescript{n}{}{\pp}$ is determined by the condition $\prescript{n}{}{P}_{33}=0$.

The assembled vector of nodal forces for the corresponding discretized weak formulation of equilibrium can then be defined by
\begin{equation}
    \prescript{n}{}{f}_a^{\alpha} (\ve \theta) = {\inte{\cl \omref}{\prescript{n}{}{P}_{aB}\bigl(\prescript{n}{}{F}_{11}(\ve X), \dots, \prescript{n}{}{F}_{22}(\ve X),\prescript{n}{}{F}_{33}(\ve X),\ve \theta\bigr)  \diffp{N^\alpha(\ve X)}{X_B} }{V_0}}
    \quad .
\end{equation}
The integration is carried out using Gaussian quadrature.
In search of a model parameterization that satisfies the weak form of balance of linear momentum, the deviation from equilibrium is quantified separately for the nodes in the interior of the domain and on force-free boundaries, and for the nodes on non-homogeneous Neumann boundaries.%
\footnote{By analogy with the corresponding forward problem, we retain the terminology of Neumann boundaries to denote the regions of the boundary where traction is prescribed.
In the present setting, the entire boundary is of Neumann type.}
For this purpose, we segment the Neumann part of the boundary,
$\domreft = \partial \varOmega_{0,\hat{\ve T}\neq \ve 0}\cup\partial \varOmega_{0,\hat{\ve T}= \ve 0}\subseteq \domref$,
with $n_\varXi \in \mathbb{N}_{>0}$ disjoint subsets, 
$\bigcup\limits_{\varXi=1}^{n_\varXi} \, \domreft^\varXi = \partial \varOmega_{0,\hat{\ve T}\neq \ve 0}$,
of non-vanishing global reaction force
$\prescript{n \!}{}{F_a}\!^{\varXi}$ for each $\varXi \in \{1,\dots,n_\varXi\}$.
Moreover, we introduce the set $\mathscr{F}=\{\alpha |\ve X^\alpha \in \cl \omref \setminus \partial \varOmega_{0,\hat{\ve T}\neq \ve 0}\}$ of free nodes that do not lie on a boundary with applied global force.
Therewith, we define the loss terms
\begin{equation}
    \Ldom (\ve \theta)= \frac{1}{C_{\!\mathscr{F}} n_\mathrm{T}} \sum_{n=1}^{n_\mathrm{T}} \sum_{\alpha \in \mathscr{F}} \prescript{n\!}{}{f}_a^{\alpha}(\ve \theta) \prescript{n\!}{}{f}_a^{\alpha}(\ve \theta)\ ,
\end{equation}
which measures the overall deviation from equilibrium for nodes that are either in the interior of the domain or on homogeneous Neumann boundaries, and
\begin{equation}
    \Lbound (\ve \theta)= \frac{1}{C_{\!\mathscr{L}} n_\mathrm{T}} \sum_{n=1}^{n_\mathrm{T}} \sum_{\varXi =1}^{n_\varXi}
    \biggl( \prescript{n \!}{}{F_a}\!^{ \varXi} - \sum_{\{\alpha |\ve X^\alpha \in \domreft^\varXi\}} \prescript{n \!}{}{f}^{\alpha}_a (\ve \theta) \biggr)
    \biggl( \prescript{n \!}{}{F_a}\!^{ \varXi} - \sum_{\{\beta |\ve X^\beta \in \domreft^\varXi\}} \prescript{n \!}{}{f}^{\beta}_a (\ve \theta) \biggr) \ ,
\end{equation}
for all nodes that lie on a boundary with applied load.
For numerical purposes, both loss contributions are normalized with suited coefficients $C_{\!\mathscr{F}}= \max_{n,e,\alpha}[\prescript{n}{}{f}^{e,\alpha}_a \prescript{n}{}{f}^{e,\alpha}_a]$ (no summation over indices ${n,e,\alpha}$) and $C_{\!\mathscr{L}}=\max_{n,\varXi}[ \prescript{n \!}{}{F_a}\!^{ \varXi} \prescript{n \!}{}{F_a}\!^{ \varXi}]$ (no summation over indices ${n, \varXi}$), respectively, where
$\prescript{n}{}{f}^{e,\alpha}_a = \int_{\cl \omrefe}{\prescript{n}{}{P}_{aB} \, \partial N^\alpha/\partial X_B \,}{\mathrm{d}V_0}$
are the element-wise nodal forces.
The model parameters $\ve \theta$ can then be identified through constrained minimization 
\begin{align}
	\hat{\ve \theta} = \underset{\ve \theta\in\mathscr C}{\arg\min} \; 	w_{\!\mathscr{F}} \Ldom(\ve \theta) +
	w_{\!\mathscr{L}} \Lbound(\ve \theta) \
	\; .
\end{align}
making use of, e.g., the SLSQP (Sequential Least Squares Programming) optimizer. For the details on the set of admissible model parameters~$\mathscr{C}$, see \cite{dammass2025b}. 
Though balancing the influence of the losses by appropriate weighting may be relevant in some cases, the choice of $w_{\!\mathscr{F}}=w_{\!\mathscr{L}}=1$ has been found to be suited for all examples throughout this work.

\section{Requirements for a proper equilibrium-gap calibration: specimen thickness and surface resolution}
In what follows, the validity of the plane stress assumption and the necessity for high surface resolution of the displacement measurement are investigated based on synthetic experiments.

\subsection{Generation of synthetic data}

\subsubsection{Ground truth constitutive model}

\begin{figure}[h]
    \centering
    \includegraphics[width=\linewidth]{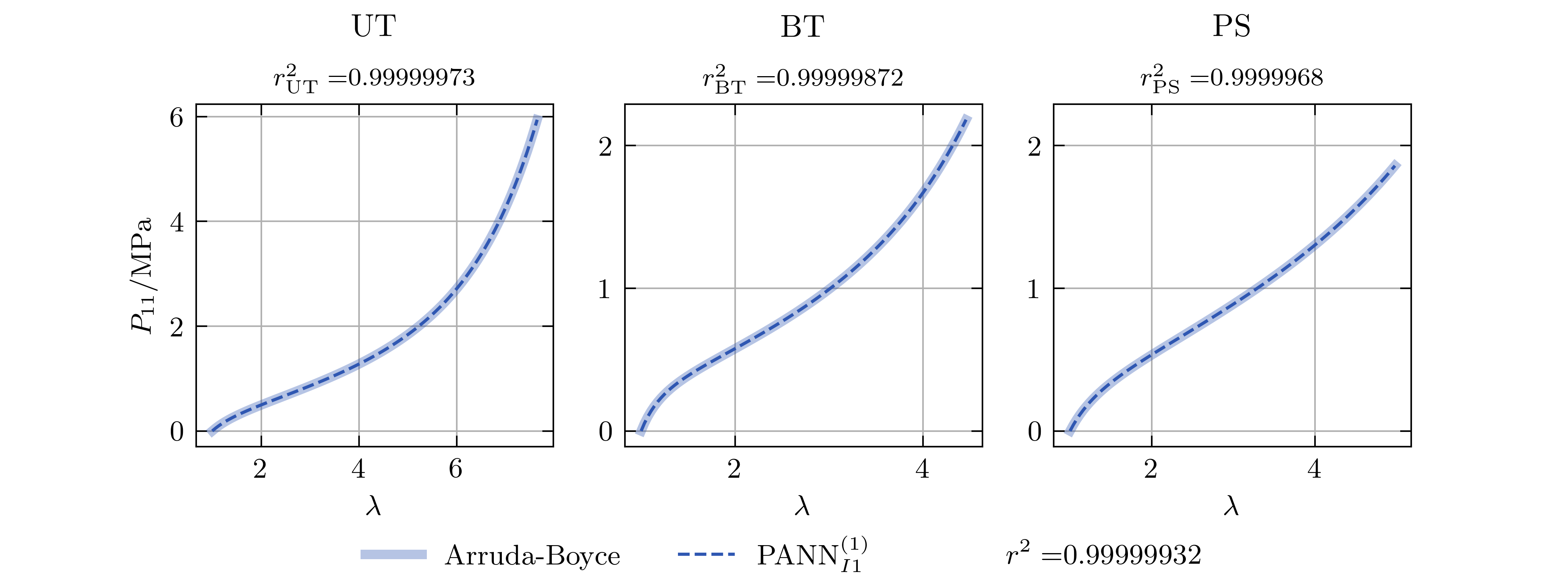}
    \caption{Supervised calibration of a PANN on the Arruda-Boyce model. The underlying neural network consists of a single hidden layer containing only one neuron and the associated skip connection.
    }
    \label{fig:supervised_PANN}
\end{figure}

For the purpose of examining the plane stress assumption and the requirement for high surface resolution, we adopt the Arruda-Boyce model in its invariant formulation \cite{arruda1993,steinmann2012} as a reference.
Rather than introducing an ad-hoc parameterization, we rely on the calibration presented by \cite{steinmann2012}, which is based on Treloar’s well-known experimental data for natural rubber~\cite{treloar1944}, probing uniaxial tension (UT), $\te F^\mathrm{UT} = \diag(\lambda,\lambda^{-1/2},\lambda^{-1/2})$, equi-biaxial tension (BT), $\te F^\mathrm{BT} = \diag(\lambda,\lambda,\lambda^{-2})$ and pure shear (PS), $\te F^\mathrm{PS} = \diag(\lambda,1,\lambda^{-1})$.
The isochoric portion of the free energy
\begin{equation}
    \psii{}^\mathrm{AB} = \mu \sum_{k=1}^5\frac{c_k}{N^{k-1}}\left( \overline{I}_1\!\!{}^k -3^k\right)
\end{equation}
is defined through the parameters $\mu = 0.2698 \ \mathrm{MPa}, \ N=21.49$ and coefficients $c_k$ related to polymer chain statistics, cf. Tab.~\ref{tab:ArrudaBoyce_params}.
\begin{table}[h]
    \caption{Parameters for the Arruda-Boyce model, taken from \cite{steinmann2012}.}
    \label{tab:ArrudaBoyce_params}
    \centering
    \begin{tabular}{lccccc}
    \toprule
        $k$& $1$& $2$ & $3$ & $4$ & $5$\\
          \midrule
         $c_k$ & $\frac{1}{2}$& $\frac{1}{20}$ & $\frac{11}{1050}$ & $\frac{19}{7000}$ & $\frac{519}{673750}$ \\
          \bottomrule
    \end{tabular}
\end{table}
Moreover, considering the deformation range covered by Treloar’s experimental data, we find that a PANN that takes only $\ici$ as input and contains a single hidden layer with one neuron and the associated skip-connection
is sufficient to approximate the Arruda-Boyce model in an excellent manner, as shown in Fig.~\ref{fig:supervised_PANN}.%
\footnote{It is not surprising that a model that depends only on $\ici$ can be represented by a PANN, that also relies solely on the first deformation invariant. The described neural network architecture thus consists of only five parameters. }
In this setup, supervised Sobolev training is performed using standard constrained optimization on the displayed loadings.
The agreement between the stress response $\te P^\mathrm{PANN} $ of the calibrated PANN and the response $\te P^\mathrm{AB}$ of the ground truth model is quantified through the coefficient of determination
\begin{equation}
    r^2 = 1 - \frac{\sum_{n=1}^{n_\mathrm{T}}
    \left( \prescript{n}{}{P}_{11}^\mathrm{AB} - \prescript{n}{}{P}_{11}^\mathrm{PANN} \right)^2}
         {\sum_{n=1}^{n_\mathrm{T}} \left( \prescript{n}{}{P}_{11}^\mathrm{AB} - \hat P_{11} \right)^2},
    \qquad \text{with} \qquad
    \hat P_{11} = \frac{1}{n_\mathrm{T}} \sum_{n=1}^{n_\mathrm{T}} \prescript{n}{}{P}_{11}^\mathrm{AB},
    \label{eq:r2_P}
\end{equation}
evaluated over all $n_\mathrm{T}$ time increments.
Accordingly, all subsequent studies based on synthetically generated data seek to calibrate this PANN with a fixed architecture within the EGM.

\subsubsection{Finite element simulations}
The following studies are carried out on a cross-shaped specimen with ellipsoidal perforations under biaxial tension, cf.~App.~\ref{app:bt_cross}.
Two different finite element settings are employed, both within a total Lagrangian framework.
The first is a two-dimensional plane stress formulation with biquadratic quadrilateral elements, used to study the influence of the surface resolution, as employed in, e.g.,~\cite{friedrichs2026,pascon2019}.
Here, the boundary value problem is solved once on a sufficiently fine mesh, and the resolution is subsequently reduced by downsampling the mesh nodes in a post-processing step.
The second is a fully three-dimensional model, which is implemented in \textit{FEniCSx}~\cite{baratta2023,scroggs2022,scroggs2022a}.
Thereby, we follow the perturbed Lagrangian approach to incompressibility, see~\cite[Sect.~10.2]{wriggers2008} and our previous work~\cite{dammass2025a}, and consider hexahedral \textit{Q2-P1} elements in order to avoid volumetric locking.
Since the 3D model resolves the stress variation across the thickness, it permits the influence of the specimen thickness to be investigated.
Convergence of the 3D computation has been verified, with regard to both the regularization of the incompressibility constraint and the through-thickness discretization.
Displacements of $u_\mathrm{max}=60$ mm are applied at each of the specimen arms over 10 increments, see~App.~\ref{app:3d_defo}.
For both settings, noise is superimposed on the displacement field and the global reaction forces following~\cite{linden2025}, cf.~App.~\ref{app:noise} for details.

\subsection{Validity of plane stress assumption for different specimen thicknesses}
\begin{figure}[h!]
    \centering
    \includegraphics[width=1\linewidth]{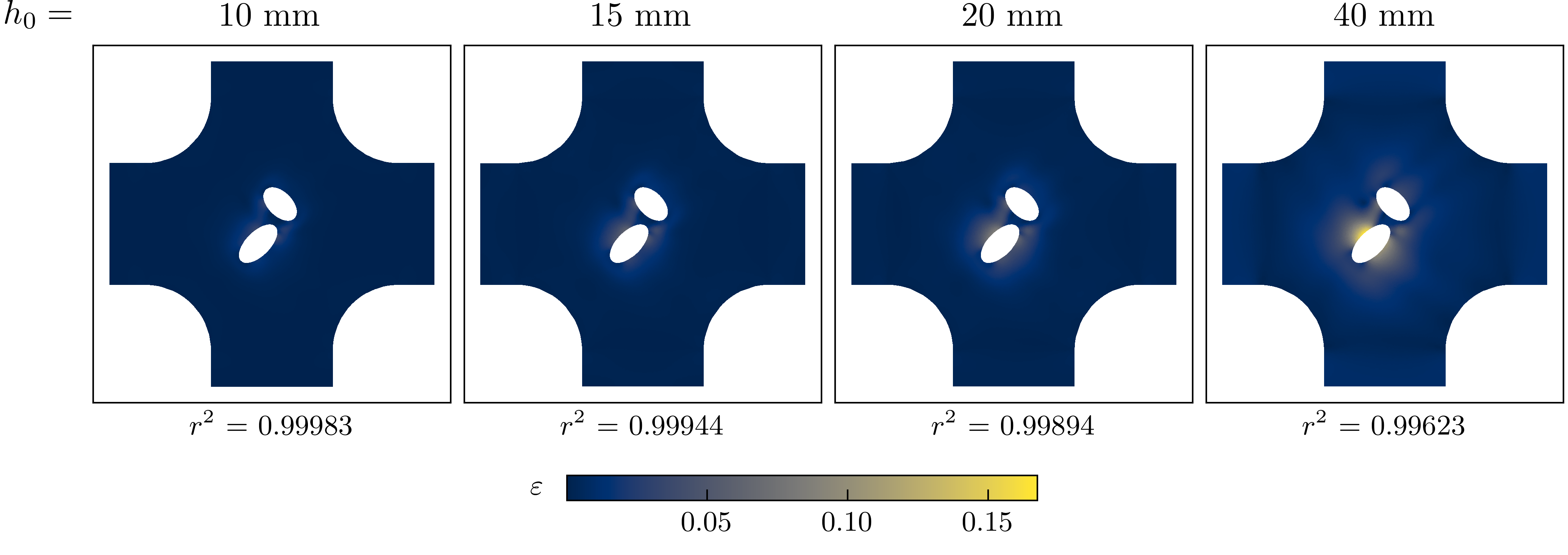}
    \caption{
    Deviation of the in-plane deformation obtained from the three-dimensional simulations from those of the ideal plane stress solution, for varying specimen thickness. 
    The deviations are most pronounced around the perforations and grow with increasing thickness.
    For dimensional reference, the semi-minor axis of the ellipses measures $3.75$ mm, cf.~Fig.~\ref{fig:BTcross_specimen_dimensions}.
    }
    \label{fig:planestress_C_diff}
\end{figure}

\begin{figure}[htb!]
    \centering
    \includegraphics[width=1\linewidth]{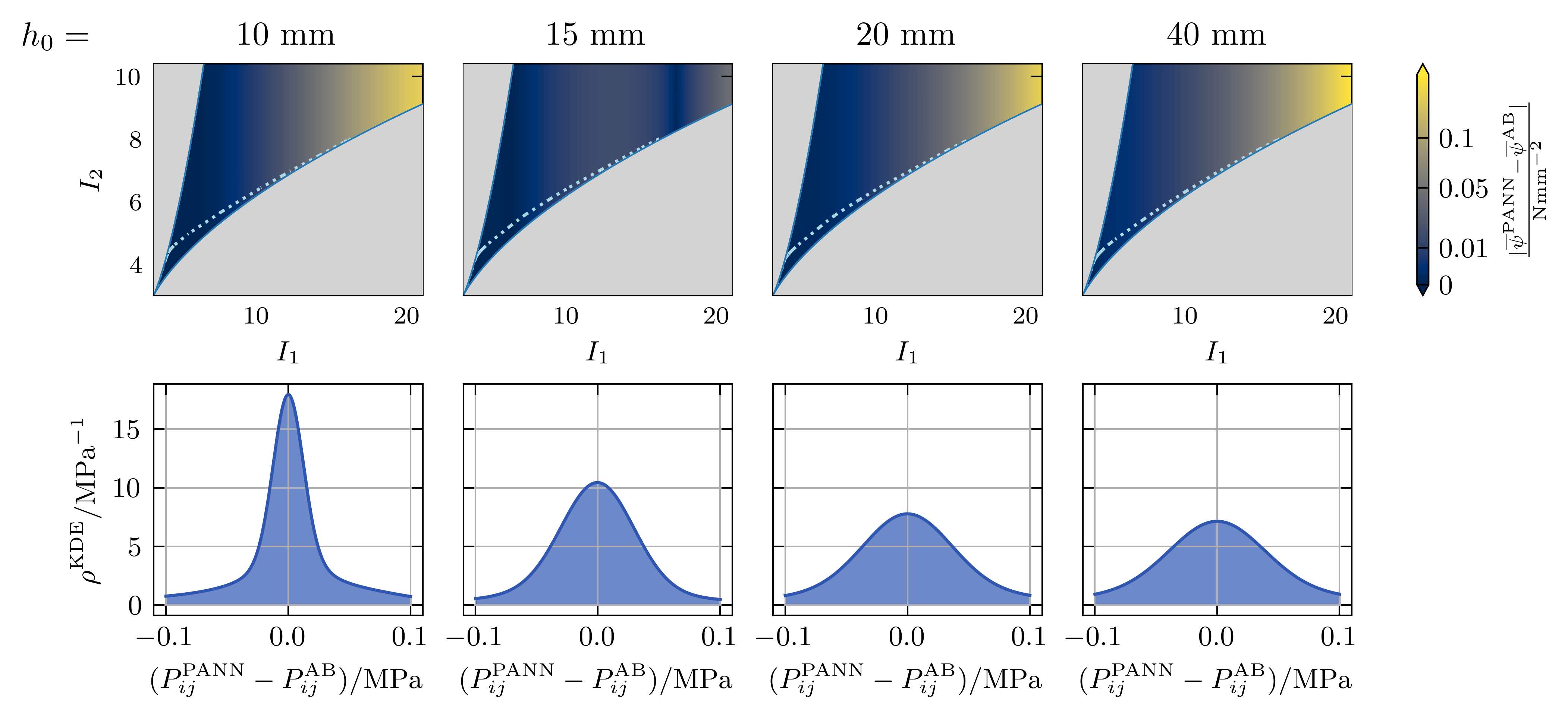}
    \caption{PANN calibrated using the noised in-plane surface displacement of progressively thicker specimens (column-wise, from left to right).
            First row: absolute deviation of the calibrated energy $\overline{\psi}{}^\mathrm{PANN}$ from the ground truth energy $\overline{\psi}{}^\mathrm{AB}$, depicted over the invariant plane.
            The convex hull of $(I_1,I_2)$-states encountered in the calibration data is indicated as a dotted line.
            Second row: Kernel density estimates $\rho^\mathrm{KDE}$ indicate the signed error distributions of the predicted stress components.
            }
    \label{fig:thickness_energy_kde}
\end{figure}

When specimen thickness is no longer small compared to the characteristic in-plane dimensions, non-negligible through-thickness stresses may arise that render the plane stress assumption a crude approximation.
In turn, this also substantially affects the in-plane deformation field on the surface of the sample, which enters the EGM.
In what follows, we investigate the consequences in the model parameterization for specimens with initial thickness $h_0 \in \{10, 15, 20, 40\}$ mm.
To quantify the deviation between the surface deformations $C_{AB}^\mathrm{3D}$ of the full 3D simulations and those from ideal plane stress computations $C_{AB}^\mathrm{2D}$, we consider a deviation measure
\begin{equation}
\varepsilon = \sqrt{\frac{(C_{AB}^{\mathrm{2D}} - C_{AB}^{\mathrm{3D}})(C_{AB}^{\mathrm{2D}} - C_{AB}^{\mathrm{3D}})}{C_{CD}^{\mathrm{2D}}C_{CD}^{\mathrm{2D}}}}
\end{equation}
depicted in Fig.~\ref{fig:planestress_C_diff} over the reference domain.
Although a thickness of $h_0=10$ mm is already of the same order as the smallest characteristic in-plane length of the considered specimen, namely the semi-minor axis of the ellipses measuring $3.75$ mm, cf.~App.~\ref{app:bt_cross}, severe deviations in the deformation field do not occur until $h_0=40$ mm.
These discrepancies, however, are located right around the holes, where the highest levels of overall deformation occur. 
Accordingly, the model parameterization is influenced rather strongly at these deformation levels and beyond, compared to the model's response for lower magnitudes of deformation.
This is most apparent in the difference between the identified amount of strain energy and the ground truth, see~Fig.~\ref{fig:thickness_energy_kde}.
Note that the former is calibrated using the noisy in-plane displacements and reaction forces from the 3D simulations, cf.~App.~\ref{app:noise}.
To investigate the stress errors across the many deformation states within the calibration data, we rely on a kernel density estimate $\rho^\mathrm{KDE}$~\cite{scott1992}.
As can be seen from Fig.~\ref{fig:thickness_energy_kde}, the errors in the predicted stress components become already significantly more pronounced at $h_0 = 15\,\mathrm{mm}$.
Overall, for the considered specimen, plane stress appears to remain an appropriate approximation up to a thickness of $h_0=10\, \mathrm{mm}$, which might be unexpected, compared with the remaining specimen dimensions.

\subsection{The role of resolution}
\label{sec:surface_resolution}
Optical full-field measurement techniques must resolve the surface motion finely enough to capture the gradients present in the deformation field. 
This requirement is particularly pronounced for the inhomogeneous specimen geometries considered here, and it is sharpened further by the EGM itself, which evaluates the equilibrium residual based on derivatives of the measured displacements.
In practice, the attainable resolution is limited by the experimental setup and the necessary size of the pixel subset for correlation, which in turn is dictated by the speckle feature size, among other influences.%
\footnote{A high point density also permits the displacement field to be resolved closer to the specimen edges, which is relevant for the boundary terms entering the EGM, among other aspects, see also Footnote~\ref{ftnt:ROI_ETH_specs}. 
This aspect is not pursued in the present study.}
Given that the displacement field can thus not be captured up to arbitrarily fine resolution, 
the question is how severe the consequences of a coarse field measurement actually are for the parameterization. 
Put differently: how coarse is too coarse, and how fine is in fact necessary? 

\begin{figure}[tbp!]
    \centering
    \includegraphics[width=1\linewidth, trim={0 0cm 0 1.2cm},clip]{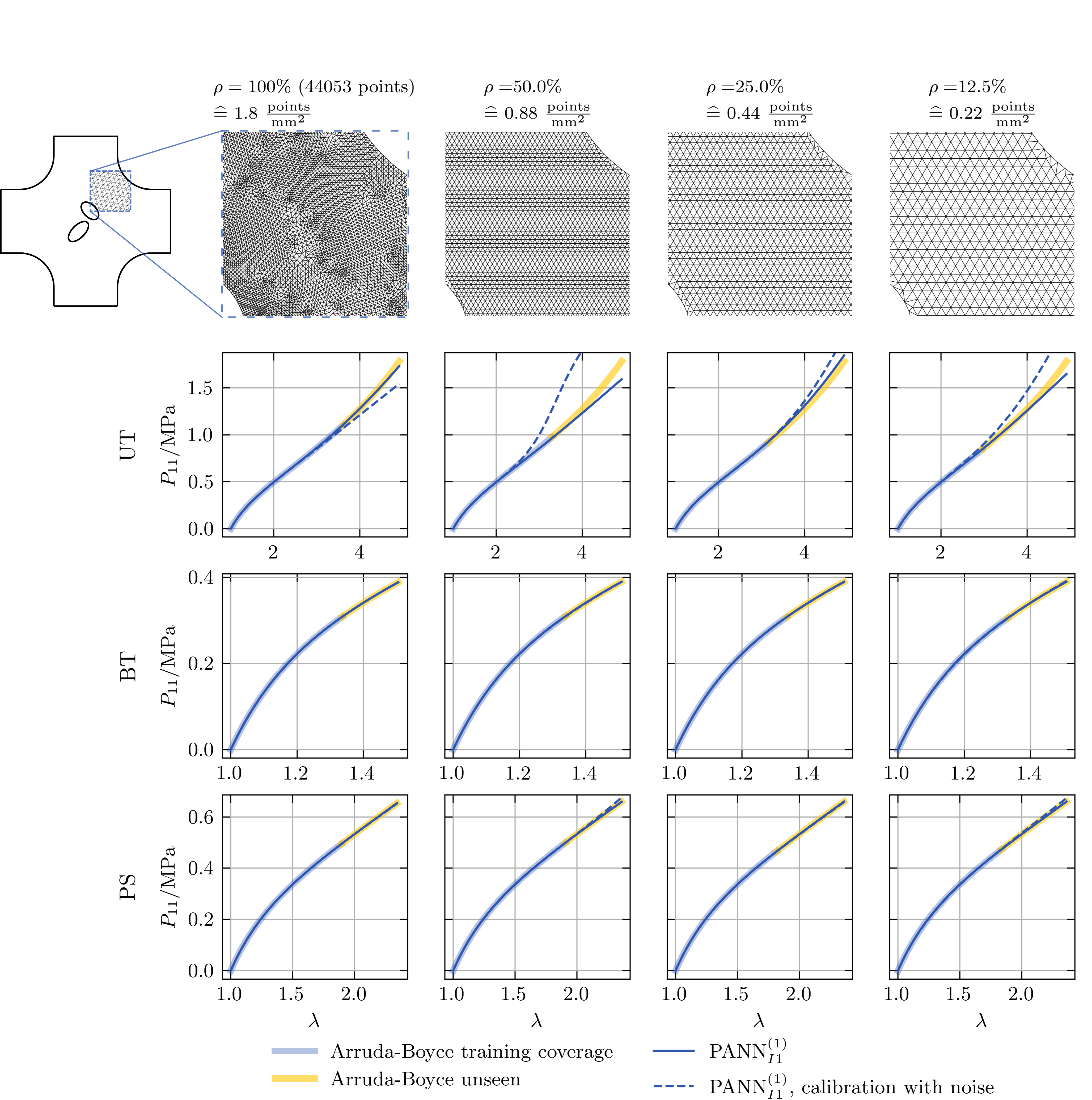}
\caption{PANNs calibrated on progressively coarser surface resolutions (column-wise, left to right). 
First row: meshes retaining a fraction $\rho$ of the initial nodes, where $\rho = 100\,\%$ corresponds to a triangulation of the nodes of the quadrilateral FE mesh used for the forward problem. 
Second to fourth row: UT, BT and PS predictions of the calibrated models, each with and without noise applied to the calibration data.
Deformation states contained within the calibration data, as well as those corresponding to extrapolation, are indicated.}
    \label{fig:mesh_resolution_study}
\end{figure}

In an attempt to provide an answer to this, we solve the boundary value problem on a sufficiently fine mesh in the 2D plane stress setting.
Thereafter, the surface resolution is subsequently reduced by 
projection of the nodal displacements onto coarser meshes, retaining a percentage $\rho \in \{100\%, 50\%, 25\%, 12.5\%\}$ of the initial nodes, as shown in Fig.~\ref{fig:mesh_resolution_study}, to which again appropriate noise is added, cf.~App.~\ref{app:noise}.
For reference, optical full-field measurements typically achieve $\sim 1\,\frac{\mathrm{point}}{\mathrm{mm}^2}$ \cite{abbasi2026a}.
Obviously, coarsening the surface resolution shortens the range of deformation encountered during calibration, as strain concentrations cannot be resolved sufficiently.
Within the deformation range encountered during calibration, however, calibration remains reliable for any $\rho$ considered, independent of the presence of noise.
Extrapolation beyond this range happens to be rather strongly affected by noisy training data, most markedly under UT, whereas
BT and PS are predicted rather accurately throughout.
So in the end, the question regarding the necessity of the surface resolution is shifted towards the question of what levels of deformation need to be covered during calibration, to ensure robust model predictions, even more so when substantial noise is present.

\section{Calibration on experimental data: specimen geometry, invariants and network size}

In what follows, we demonstrate the applicability of PANN models within the EGM using experimental full-field displacement data.
We assess calibration performance on real data, investigate robustness and extrapolation capabilities based on specimen geometry, and study the roles of input invariants and network architecture size.
The experimental data employed in this study are taken from~\cite{abbasi2026a} and consist of full-field displacement measurements obtained via digital image correlation (DIC) on several geometries of rubber specimens (NR-40) subjected to quasi-static axial loading.
The considered geometries consist of classical specimens for uniaxial tension (UT) and pure shear (PS),%
\footnote{UT and PS denote both the load cases and the corresponding specimen geometries, which contain slight deviations from the ideal homogeneous states. 
The intended meaning follows from the context.}
as well as specimens with inhomogeneous geometry, referred to as TTa, \dots, TTf, where holes are introduced in the domain with the purpose of achieving a wide range of multiaxial deformation states within a single specimen, see Tab.~\ref{tab:eth_specs}.

\begin{table}[h]
    \centering
    \caption{Test specimens under tension, for which global force and full-field displacement measurements are available from the experiments described in \cite{abbasi2026a}.}
    \begin{tabular}{c c c c c c c c}
        \toprule
        \small  
         UT & PS & TTa & TTb & TTc & TTd & TTe & TTf\\
         \midrule
        \includegraphics[height=0.1\textwidth,decodearray={-1.5 1 -1.5 1 -1.5 1}]{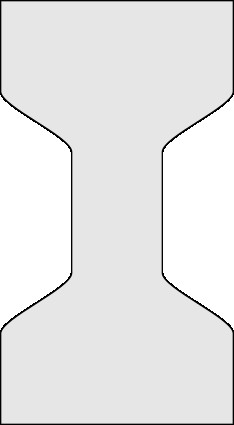} & 
        \raisebox{12pt}{\includegraphics[height=0.05\textwidth,decodearray={-1.5 1 -1.5 1 -1.5 1}]{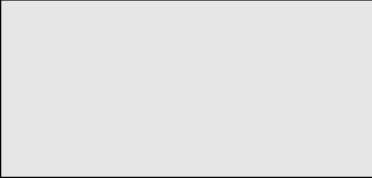}} & 
        \includegraphics[height=0.1\textwidth,decodearray={-1.5 1 -1.5 1 -1.5 1}]{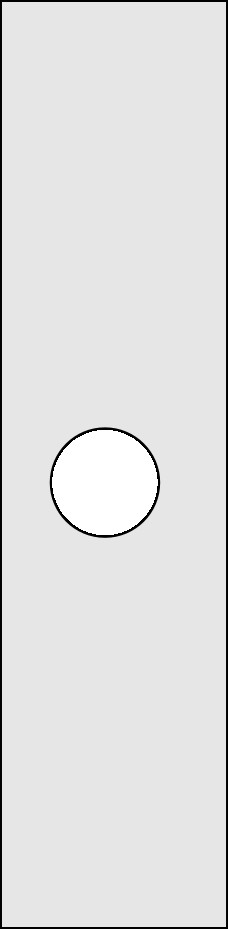}& 
        \includegraphics[height=0.1\textwidth,decodearray={-1.5 1 -1.5 1 -1.5 1}]{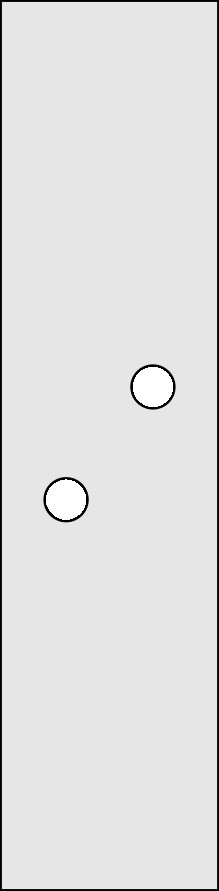}& 
        \includegraphics[height=0.1\textwidth,decodearray={-1.5 1 -1.5 1 -1.5 1}]{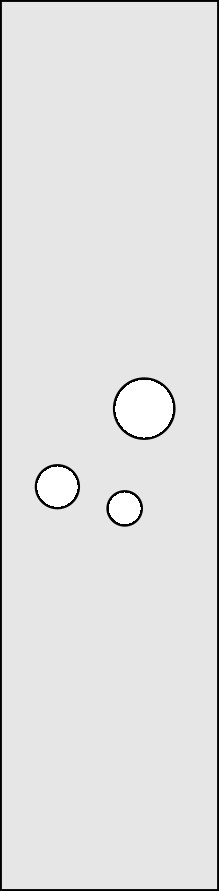}& 
        \includegraphics[height=0.1\textwidth,decodearray={-1.5 1 -1.5 1 -1.5 1}]{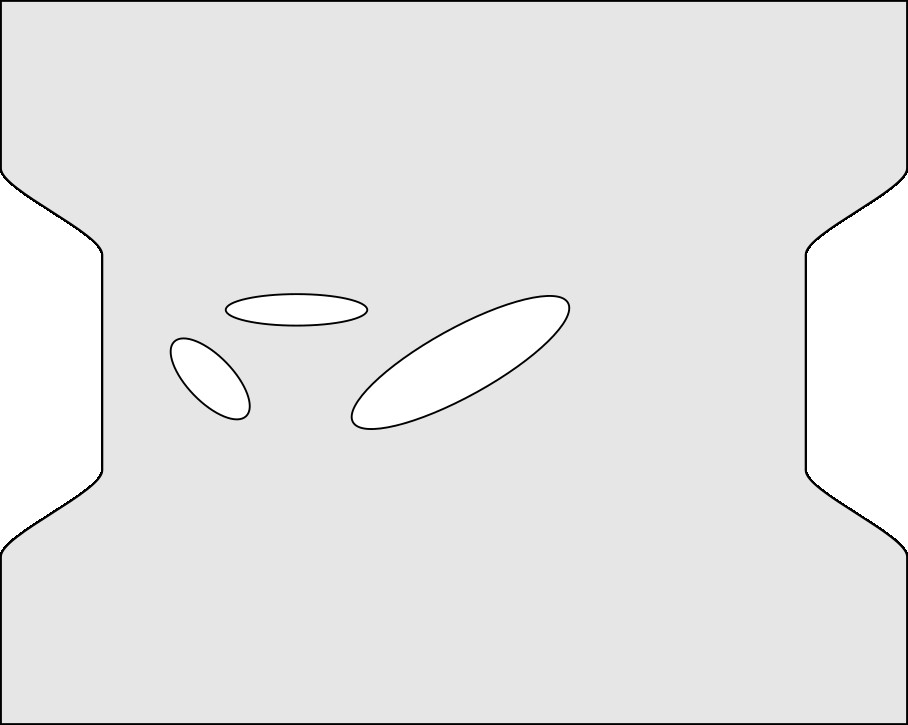}& 
        \includegraphics[height=0.1\textwidth,decodearray={-1.5 1 -1.5 1 -1.5 1}]{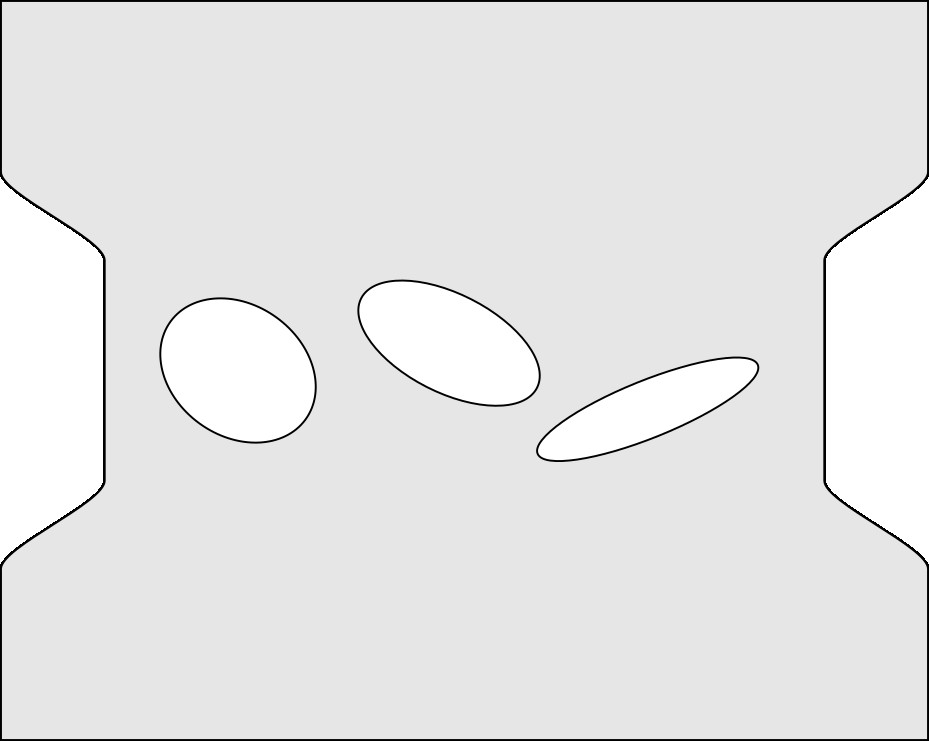}&
        \includegraphics[height=0.1\textwidth,decodearray={-1.5 1 -1.5 1 -1.5 1}]{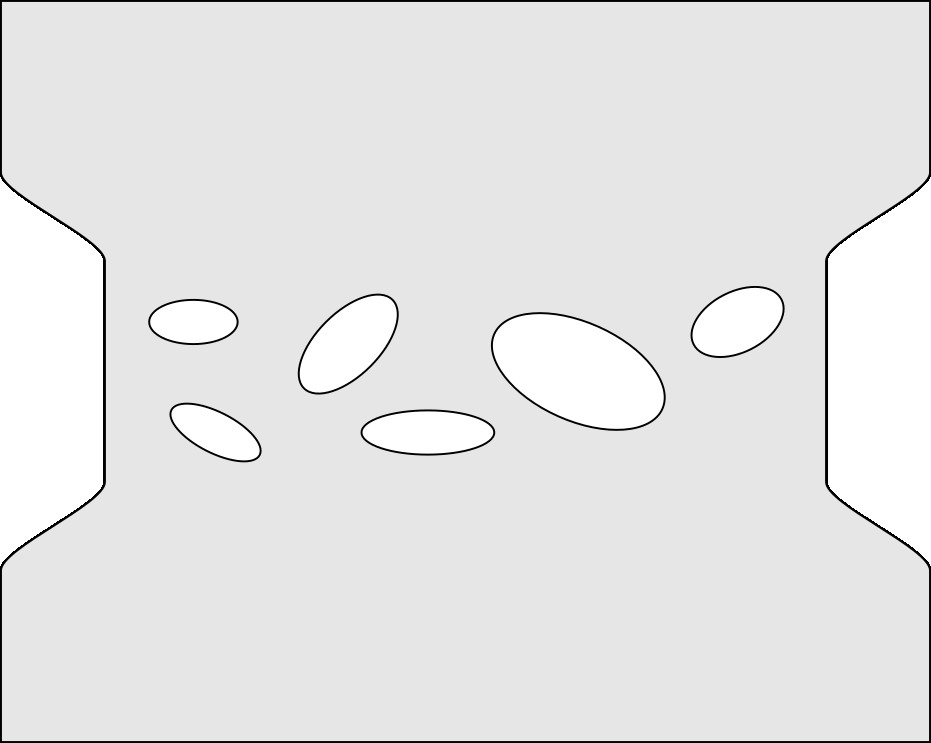}\\
        \bottomrule
    \end{tabular}
    \label{tab:eth_specs}
\end{table}
Furthermore, various PANN architectures with different network sizes and different combinations of input invariants are calibrated on each specimen and evaluated on the load cases UT, PS, and BT to assess the extrapolation capabilities and robustness of each parameterization, where reference stress--stretch relations for UT and PS are determined under the classical assumption of homogeneous deformation \cite{abbasi2026a}.%
\footnote{\label{ftnt:ROI_ETH_specs}A crucial yet often disregarded aspect of full-field calibration methods such as the EGM is that the displacement fields obtained from DIC cannot be measured up to the specimen boundaries. 
Consequently, specimen dimensions, e.g., the width, are underestimated when only the region of interest is considered, and stress estimates under nominally homogeneous deformation states become prone to error. 
In particular, the stress response for, e.g., UT in \cite{abbasi2026a} is computed using the width of the region of interest ($25$ mm) rather than the actual specimen width ($30$ mm).}
Specifically, network architectures with $\mathcal{N}\in\{(1),(8),(2, 2), (4, 4), (16,16)\}$ have been examined within the scope of this work, where $\mathcal{N}=(1)$ denotes a single hidden layer with one neuron only, whereas $\mathcal{N}=(16,16)$ denotes two hidden layers with $16$ neurons each, etc.
In what follows, we put an emphasis on the PANN architecture with $\mathcal{N}=(4,4)$, which was found to provide sufficient flexibility for describing the underlying constitutive behavior while remaining compact.

\subsection{The role of specimen geometry}

\begin{figure}
    \centering
    \includegraphics[width=1\linewidth, trim={0 1cm 0 0cm},clip]{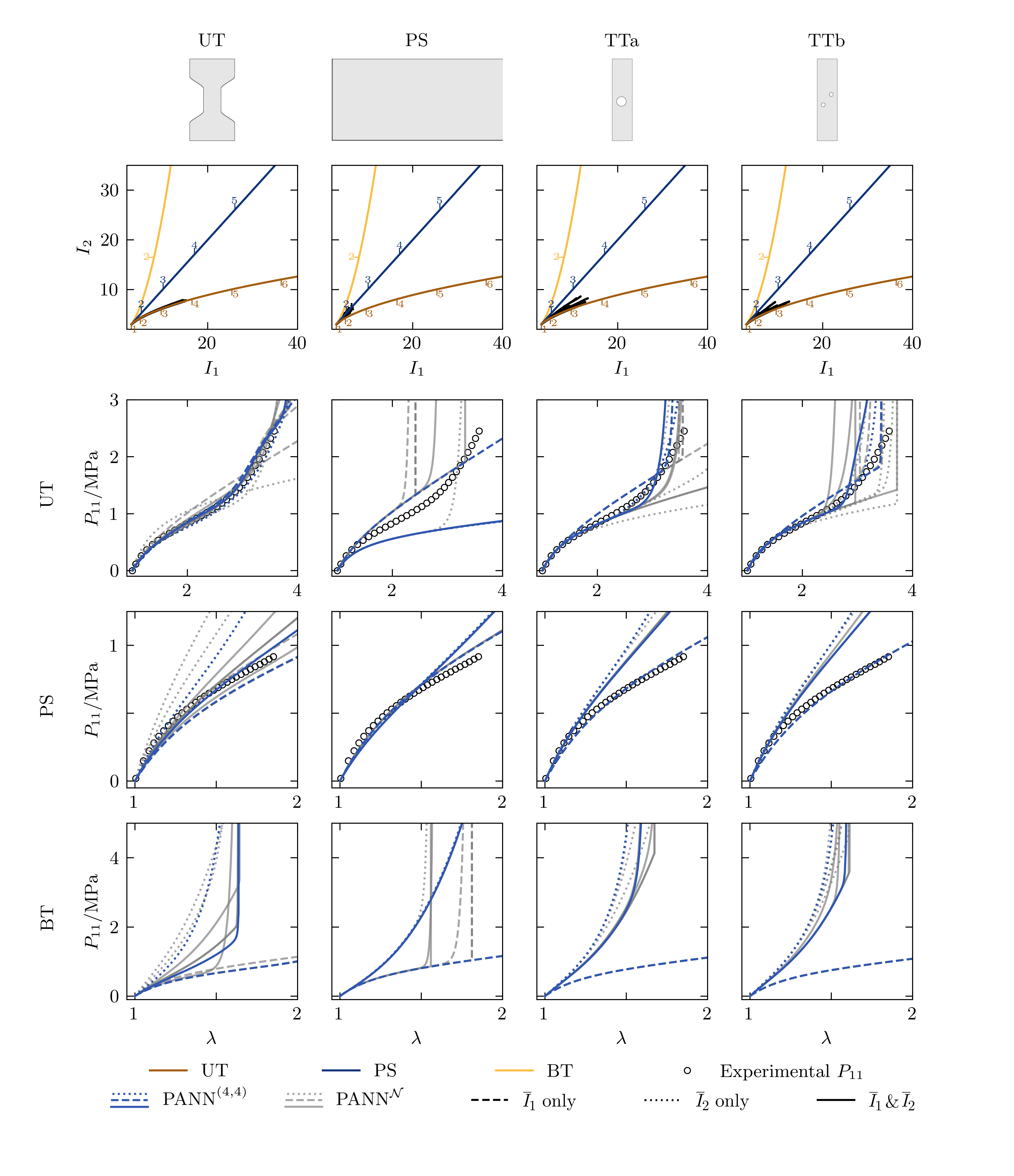}
    \caption{PANNs calibrated on specimens UT, PS, TTa and TTb (column-wise), using either only $\ici$, only $\iici$ or both. 
    First row: deformation states $(\ici, \iici)$ induced by the respective specimen geometry according to the experimental measurements, with the load cases UT, PS and BT indicated as lines and the associated stretches annotated.
    Second to fourth row: model predictions for homogeneous uniaxial, pure shear and equi-biaxial deformation. 
    The remaining network sizes $\mathcal{N} \in \{(1), (8), (2,2), (16,16)\}$ behave comparably.}
    \label{fig:SoloCalibSpecs_UT-TTb}
\end{figure}

\begin{figure}
    \centering
    \includegraphics[width=1\linewidth, trim={0 1cm 0 0cm},clip]{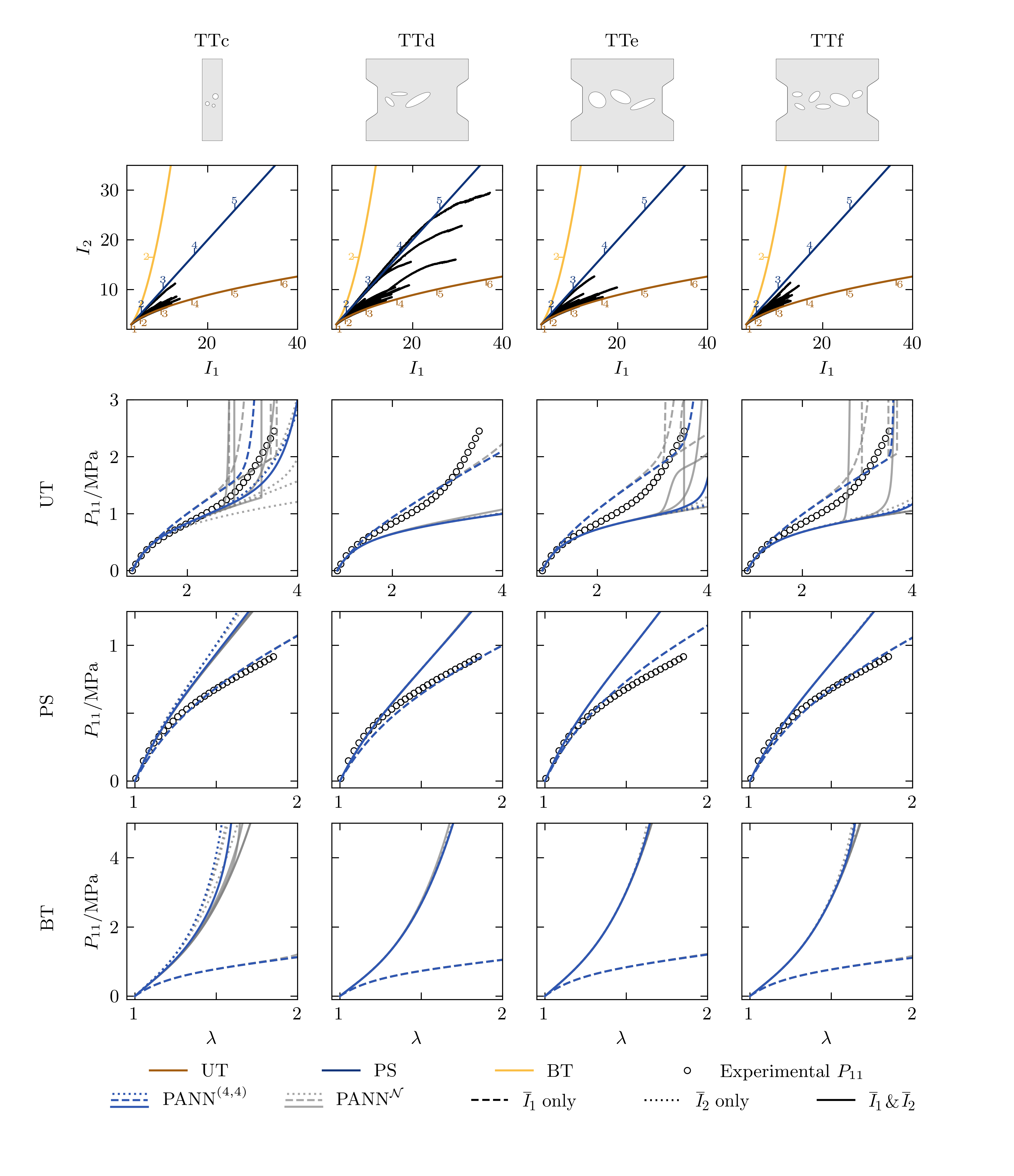}
\caption{PANNs calibrated on specimens TTc, TTd, TTe and TTf (column-wise), using either only $\ici$, only $\iici$ or both. 
    First row: deformation states $(\ici, \iici)$ induced by the respective specimen geometry according to the experimental measurements, with the load cases UT, PS and BT indicated as lines and the associated stretches annotated.
    Second to fourth row: model predictions for homogeneous uniaxial, pure shear and equi-biaxial deformation, the former two of which the true stress--stretch behavior is available for. 
    The remaining network sizes $\mathcal{N} \in \{(1), (8), (2,2), (16,16)\}$ behave comparably.}
    \label{fig:SoloCalibSpecs_TTc-TTf}
\end{figure}

Since each specimen exhibits a different displacement field, every experiment covers a different portion within the set of admissible deformation states. 
Under incompressibility, this set can be characterized through the invariant tuples $(\ici, \iici) \in \rs^2$ that are attainable for some $\ciso \in \tesu \cap \mathcal{T}^2_\text{s>}$. 
Its bounds coincide with the load cases of uniaxial and equi-biaxial tension expressed in terms of the invariants, see~\cite[Theorem~2 and Corollary~3]{dammass2025b}.
The deformation states obtained for each specimen are visualized in the first row of Figs.~\ref{fig:SoloCalibSpecs_UT-TTb} and~\ref{fig:SoloCalibSpecs_TTc-TTf}.
The classical UT and PS specimens deviate only slightly from the intended homogeneous deformation and, as expected, cover only very narrow subsets of the admissible invariant set, whereas the inhomogeneous geometries achieve a considerably broader coverage, most notably specimen TTd.

The consequence of this coverage becomes apparent in the predicted UT, PS and BT responses shown in the remaining rows of Figs.~\ref{fig:SoloCalibSpecs_UT-TTb} and~\ref{fig:SoloCalibSpecs_TTc-TTf}, where two effects are observed.
First, the reference stress--stretch relations are reproduced more accurately, the wider the range of deformation states sampled by the calibration specimen. 
Second, the richer this coverage, the more closely the extrapolated responses of models with different network architectures coincide.
A wide coverage of the deformation space thus goes along with a more robust, or at least less ambiguous, extrapolation towards unseen deformation states.
Trustworthy extrapolation towards deformation levels beyond those encountered during calibration cannot, however, be guaranteed, just as for classical calibration techniques.

\begin{figure}
    \centering
    \includegraphics[width=1\linewidth, trim={0 1cm 0 0cm},clip]{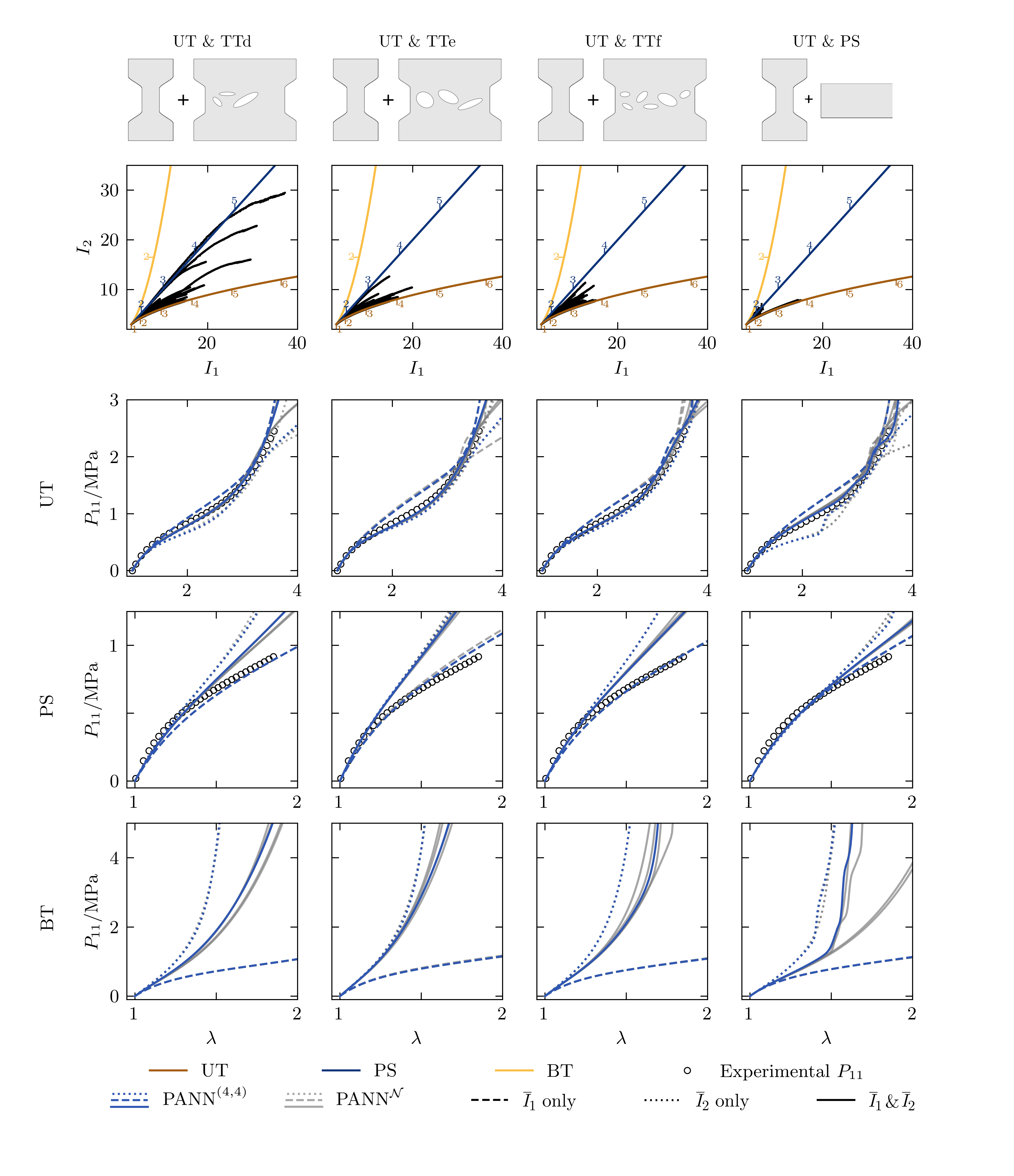}
    \caption{PANNs jointly calibrated on the UT specimen combined with TTd, TTe, TTf or PS (column-wise), using either only $\ici$, only $\iici$ or both. 
    First row: deformation states $(\ici, \iici)$ induced by the respective specimen combination according to the experimental measurements, with the load cases UT, PS and BT indicated as lines and the associated stretches annotated. 
    Second to fourth row: model predictions for homogeneous uniaxial, pure shear and equi-biaxial deformation, the former two of which the true stress--stretch behavior is available for. 
    The remaining network sizes $\mathcal{N} \in \{(1), (8), (2,2), (16,16)\}$ behave comparably.}
    \label{fig:CombiCalibSpecs}
\end{figure}

Since each complex experiment likewise samples only a portion of the set of admissible deformations, the question arises whether several specimens can be combined in a complementary manner, to enrich the coverage of deformation states encountered during calibration even further.
To investigate this, we jointly calibrate on the UT specimen combined with TTd, TTe, TTf or PS, cf.~Fig.~\ref{fig:CombiCalibSpecs}. 
The UT specimen enters every combination since no other experiment provides insight into the constitutive behavior up to comparably large
stretches.
The combined calibration indeed enriches the deformation states encountered during calibration, and both effects identified above improve accordingly.
Exploiting this enrichment, however, presupposes a sufficiently flexible constitutive ansatz.
Classical models, with an often limited number of parameters and prescribed functional form, can often not fully capture the information contained in a richer deformation state space, so that the calibration becomes limited by the model rather than by the data.
Neural-network-based formulations impose no such restriction.

\begin{figure}[h]
    \centering
    \includegraphics[width=1\linewidth, trim={0 0.7cm 0 0.7cm},clip]{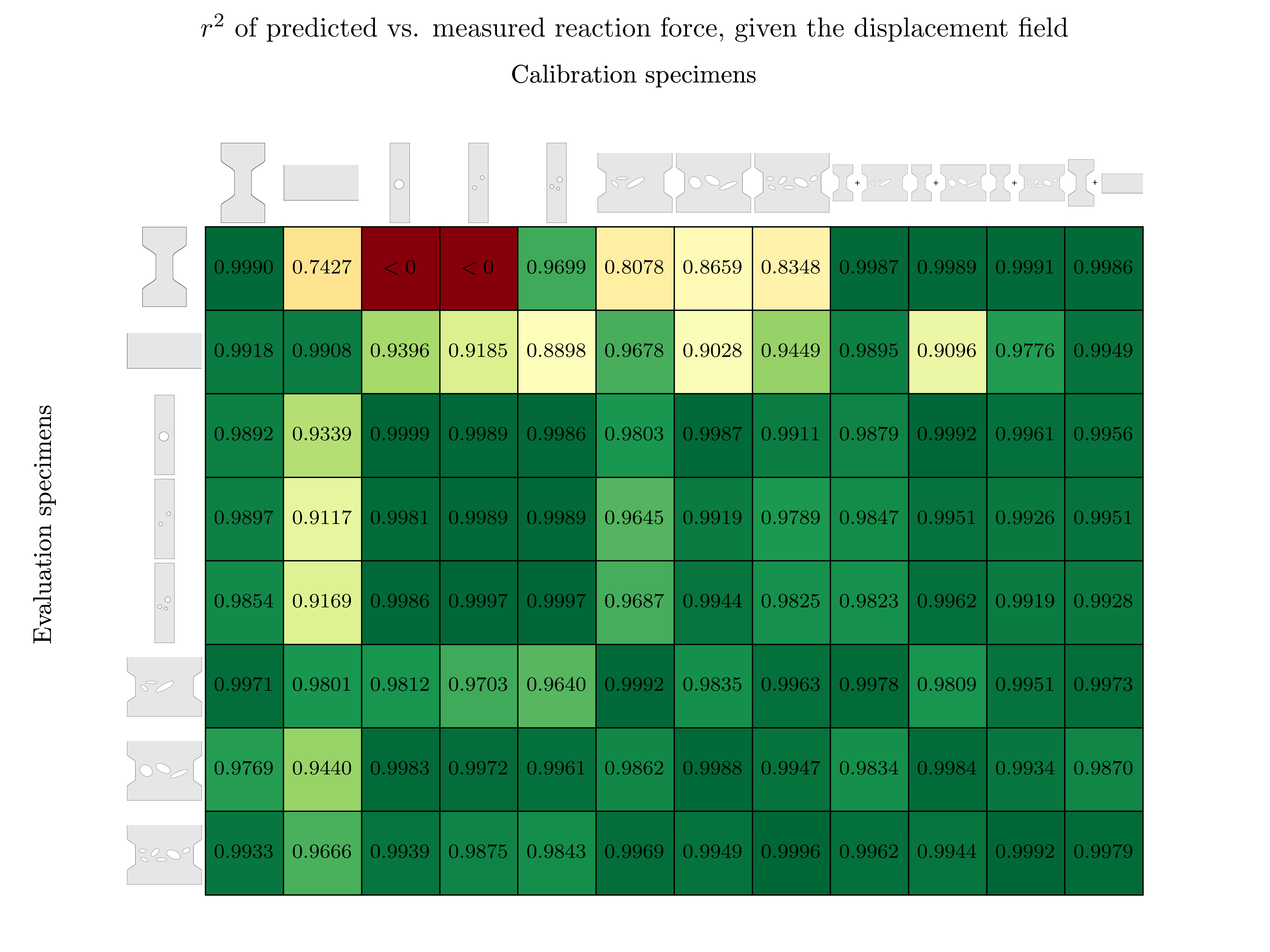}
    \caption{$r^2_\mathrm{F}$ value between predicted and measured reaction force for a PANN with $\ici$ and $\iici$ as input invariants and $\mathcal{N} = (4,4)$. 
    Columns indicate the specimen, or combination of specimens, used for calibration; rows the specimen used for evaluation, with the respective geometries depicted alongside. 
    Reaction forces are predicted under the given experimentally measured displacement field.}    
\label{fig:r2_matrix}
\end{figure}

To quantify these observations beyond the homogeneous load cases, we examine the global reaction forces predicted for a model calibrated on a given specimen, when the measured displacement field of another specimen is imposed, in analogy to the loss $\Lbound$.
The agreement between prediction and measurement is quantified through
\begin{equation}
    r^2_\mathrm{F} = 1 - \frac{
    \sum\limits_{n=1}^{n_\mathrm{T}} \sum\limits_{\varXi =1}^{n_\varXi}
    \Bigl( \prescript{n \!}{}{F_a}\!^{ \varXi} - \sum\limits_{\{\alpha |\ve X^\alpha\in \domreft^\varXi\}} \prescript{n \!}{}{f}^{\alpha}_a (\ve \theta) \Bigr)
    \Bigl( \prescript{n \!}{}{F_a}\!^{ \varXi} - \sum\limits_{\{\beta |\ve X^\beta\in \domreft^\varXi\}} \prescript{n \!}{}{f}^{\beta}_a (\ve \theta) \Bigr)
    }
    {
    \sum\limits_{n=1}^{n_\mathrm{T}} \sum\limits_{\varXi =1}^{n_\varXi}
    \Bigl( \prescript{n \!}{}{F_a}\!^{ \varXi}  - \hat{F_a}\!^{ \varXi} \Bigr)
    \Bigl( \prescript{n \!}{}{F_a}\!^{ \varXi}  - \hat{F_a}\!^{ \varXi} \Bigr)
    }
    \quad \mathrm{with} \quad
    \hat{F_a}\!^{ \varXi} = \frac{1}{n_\mathrm{T}} \sum_{n=1}^{n_\mathrm{T}} \prescript{n \!}{}{F_a}\!^{ \varXi} \point
    \label{eq:r2_F}
\end{equation}
Read column-wise, Fig.~\ref{fig:r2_matrix} shows that the combined calibration sets consistently lead to favorable predictions across all evaluation specimens. 
Calibration on the UT specimen alone likewise yields adequate predictions throughout, whereas relying exclusively on PS proves insufficient, most evidently for the UT response.
Beyond this, the design of the inhomogeneous specimens has a pronounced influence as well.
Calibration on TTa or TTb fails to reproduce the reaction force of the UT specimen altogether, while the remaining more complex geometries suffice for this purpose to a good extent, and also contain information regarding multiaxial deformation states.
Read row-wise, the two classical specimens are at the same time the most demanding evaluation cases.
The UT specimen discriminates most sharply between the calibration sets, its predictions ranging from excellent to entirely inadequate, which stems from the large stretches attained in this experiment and the correspondingly strong degree of extrapolation demanded of models calibrated elsewhere. 

For the data set under consideration, combining the UT specimen with a complex geometry, most notably TTd, essentially yields the most diverse basis for parameterization, provided the constitutive model is sufficiently flexible.
Notably, the combination of the classical UT and PS specimens, evaluated in a full-field manner, already leads to a marked improvement over single-specimen calibration, which is of practical relevance since both simple geometries are standard in many experimental laboratories.
Nevertheless, as becomes clear from Figs.~\ref{fig:SoloCalibSpecs_UT-TTb},~\ref{fig:SoloCalibSpecs_TTc-TTf}~and~\ref{fig:CombiCalibSpecs} and our previous work~\cite{dammass2025b}, reliable predictions for BT and loadings with comparable biaxiality generally require calibration data resembling multiaxial tension. These are, however, not included in the experimental data set of~\cite{abbasi2026a}.

\subsection{The role of input invariants and the network size}

While the previous study addressed the specimen side of the EGM calibration, we now turn to the model, and investigate how the choice of input invariants, only $\ici$, only $\iici$ or the pair $(\ici, \iici)$, and the size of the network affect the fitting and the extrapolation capabilities.
Figs.~\ref{fig:SoloCalibSpecs_UT-TTb},~\ref{fig:SoloCalibSpecs_TTc-TTf} and~\ref{fig:CombiCalibSpecs} reveal a distinct separation between the invariant configurations, which manifests itself almost exclusively in the extrapolation regime.
Whenever both invariants are taken into account, $\iici$ appears to dominate the response, such that
the $\iici$-only and the $\ici$\,\&\,$\iici$ models predict very similar responses under UT, PS and BT.
The $\ici$-only model, in contrast, agrees more closely with the reference UT and PS relations.
Under equi-biaxial tension, moreover, the difference is qualitative rather than merely quantitative: both configurations that include $\iici$ predict a stress that rises almost vertically already at moderate stretches, whereas only the $\ici$-only formulation remains physically plausible.
This is a direct consequence of the deformation modes covered by the calibration data, as $\iici$ grows much more rapidly under BT than under UT, cf.~\cite{dammass2025b}.
A pronounced $\iici$-dependence, identified from data that essentially comprise uniaxial and shear-dominated states, is therefore evaluated far outside the sampled range as soon as biaxial tension is approached.
Since the experimental data set of~\cite{abbasi2026a} contains no BT-like states, this deficiency is not penalized by any error measure evaluated on the available specimens.
These observations agree with~\cite{dammass2025b}, according to which multiaxial data are strictly required for the parameterization of models accounting for $\iici$, whereas $\ici$-only formulations still extrapolate in a qualitatively correct manner if such data are missing.
Combining specimens enriches the coverage of the deformation space, but does not resolve this issue.
For the joint calibrations UT\,\&\,TTd or UT\,\&\,PS, which sample a considerable range of the admissible deformation states, the additional flexibility offered by $\iici$ is indeed beneficial within the range of states actually probed.
None of the available specimens or their combinations, however, probes biaxial tension, so none of them cures the unphysical biaxial response of the $\iici$-containing models.
What is decisive for a reliable parameterization is thus not the richness of the deformation coverage as such, but whether the calibration data comprise the deformation modes that are to be predicted.
Restricting the model to $\ici$ consequently remains the safer choice whenever BT-like data are unavailable, at the cost of a reduced flexibility within the sampled range, whereas a precise calibration of models including $\iici$ presupposes calibration data that resemble multiaxial tension.

\begin{figure}[h]
    \centering
    \includegraphics[width=1\linewidth]{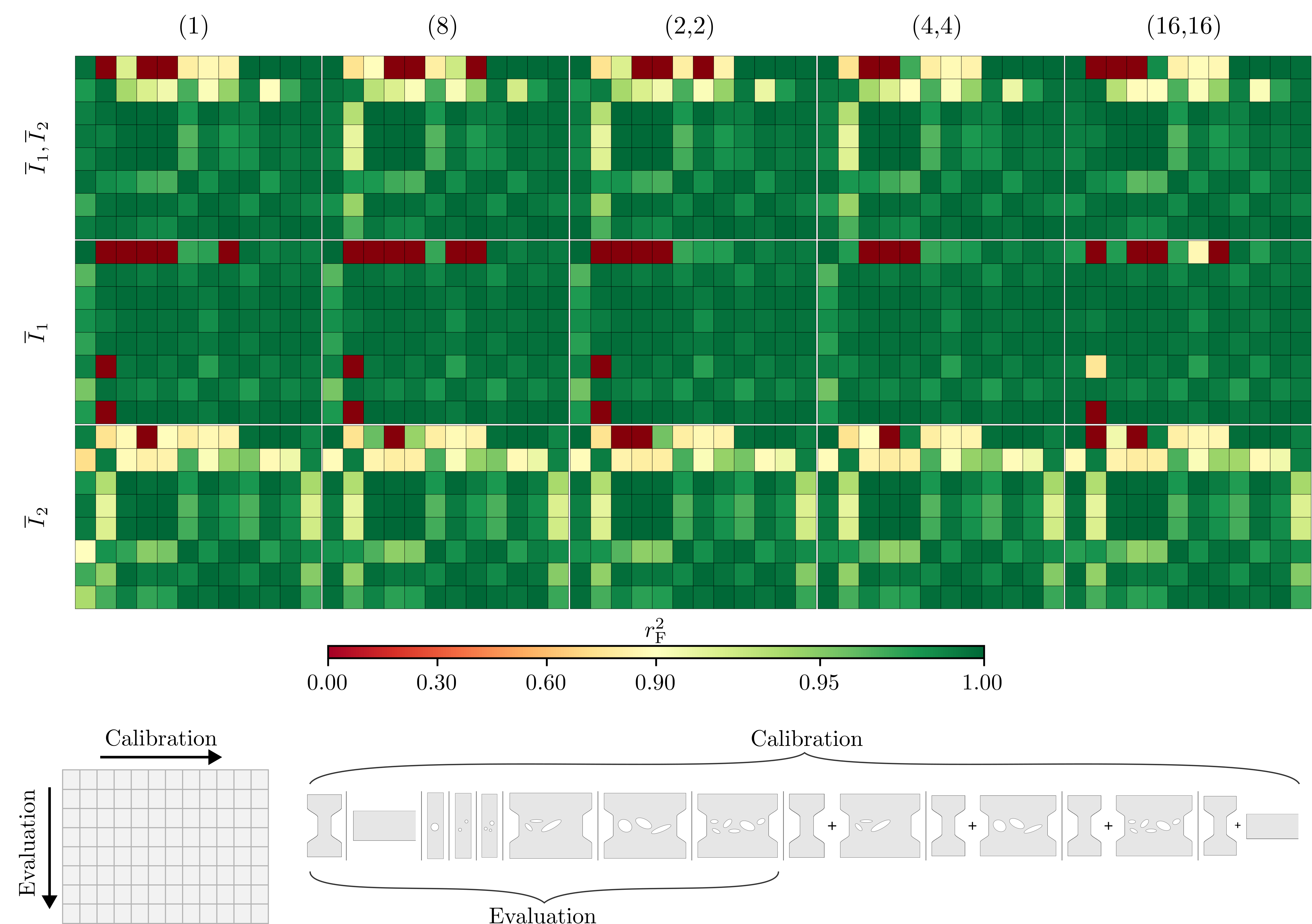}
    \caption{$r^2_\mathrm{F}$ scores of the reaction force, given the full-field displacement of the specimens, cell-wise identical to Fig.~\ref{fig:r2_matrix}. 
    Rows: varying invariants considered as input for the PANN. 
    Columns: varying number of hidden layers and neurons of the neural network. 
    }
    \label{fig:gigamatrix}
\end{figure}

To assess the extrapolation capabilities away from the homogeneous load cases, we again evaluate $r^2_\mathrm{F}$ in the array-wise manner introduced above, now for all three invariant configurations and for various network sizes, cf.~Fig.~\ref{fig:gigamatrix}.

We find that, first, a network size of $\mathcal{N} = (4,4)$ is sufficient for the underlying material, irrespective of the input invariants, while larger networks show signs of overfitting, and smaller ones show insufficient expressivity. 
Second, on the available specimens, the $\ici$\,\&\,$\iici$ models generally achieve the highest scores, whereas the $\iici$-only model behaves unreliably on unseen deformation states in different specimens.
As emphasized above, this ranking reflects only the deformation modes contained in the data, and it does not reveal the unphysical biaxial extrapolation of the $\iici$-containing formulations, since none of the specimens probes BT. 
Third, independently of both the architecture and the input invariants, a rich coverage of the deformation space during calibration improves the extrapolation capabilities of every model.
Throughout, calibration on the UT specimen alone already yields accurate predictions for the remaining geometries, presumably because uniaxial tension remains the dominant deformation state even in the inhomogeneous specimens.

\section{Conclusion}

In the present study, the calibration of physics-augmented neural networks via the equilibrium gap method is examined with respect to three practically relevant aspects: 
the limitations of the plane-stress assumption for considerable specimen thickness, 
the influence of a coarsely resolved surface displacement field on the identified model,
and the calibration on experimental full-field data for various specimen geometries. 
For the calibration on experimental data, we investigate which specimen geometry provides beneficial coverage of the deformation state space, and how the input invariants and the neural network size affect the fitting and extrapolation capabilities of the identified models.

We demonstrate that both an insufficient resolution of the surface displacements and a specimen thickness violating the plane stress assumption have the strongest influence on the parameterization for the largest deformations present in the calibration data. 
The extrapolation beyond that range degrades accordingly.
Inhomogeneous specimen geometries, and combinations thereof, provide a richer coverage of the deformation space and thus more insight into the constitutive behavior, provided the considered model is flexible enough to capture it. 
Surprisingly, combining classical specimens with approximately homogeneous displacement fields does already provide a solid basis for calibrating constitutive models.
Unless the calibration data comprise biaxial-tension-like states, however, dropping the second invariant yields considerably more reliable generalization, since formulations that include it extrapolate unphysically towards equi-biaxial tension.
A precise calibration of such models therefore requires data resembling multiaxial tension in the first place.

\section*{Acknowledgments}
Support for this work was provided by the German Research Foundation (DFG) under the grants KA 3309/9-1 (project 420422342) and KA 3309/20-1 (project 517438497).\\
The authors gratefully acknowledge Brain Riemer for the fruitful discussions.

\section*{CRediT author contribution statement}
\textit{Konrad Friedrichs:} Conceptualization, Formal analysis, Investigation, Methodology, Visualization, Software, Validation, Data Curation, Writing –- original draft, Writing –- editing.
\textit{Franz Dammaß:} Conceptualization, Formal analysis, Investigation, Methodology, Software, Writing –- editing, Supervision, Funding acquisition.
\textit{Karl A. Kalina:} Conceptualization, Formal analysis, Investigation, Methodology, Software, Writing –- editing, Supervision.
\textit{Markus Kästner:} Resources, Funding acquisition, Supervision.

\section*{Declaration of competing interest}
There is no conflict of interest to declare.

\section*{Declaration of generative AI and AI-assisted technologies in the manuscript preparation process}
During the preparation of this work the authors used Anthropic’s language models Sonnet and Opus to aid in text refinement and grammar checking. After using these tools, the authors reviewed and edited the content as needed and take full responsibility for the content of the published article.
\newpage
\appendix
\section{Appendix}
\subsection{Specimen geometry for studies on synthetic data}
\label{app:bt_cross}
\begin{figure}[h]
    \centering
    \includegraphics[height=0.6\linewidth]{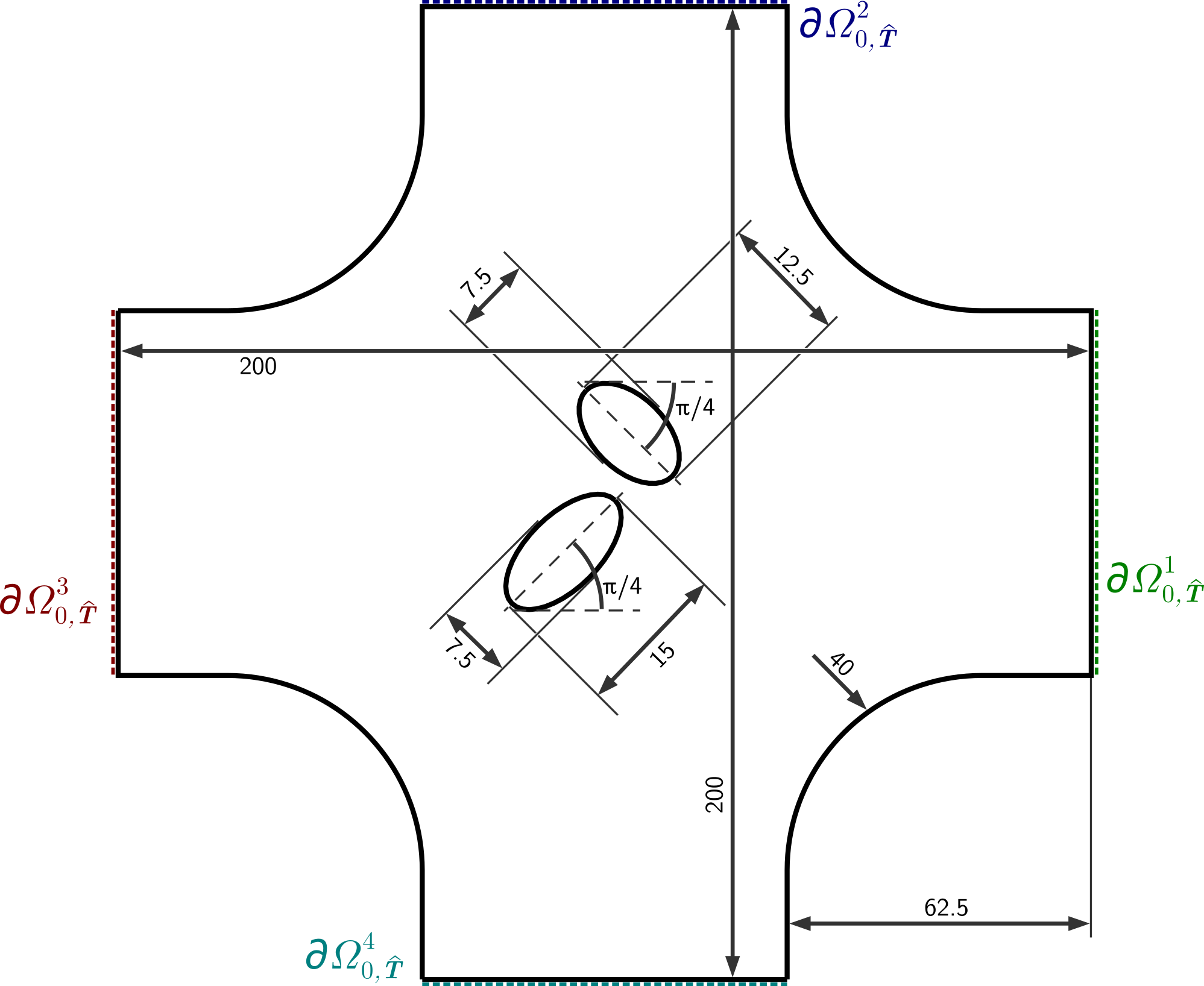}
    \caption{Drawing of the cross-shaped specimen with ellipsoidal holes considered in the synthetic data studies. Relevant dimensions and non-homogeneous Neumann boundaries are indicated. Specimen thickness is varied and depends on the specific study.}
    \label{fig:BTcross_specimen_dimensions}
\end{figure}

\subsection{3D deformation}
\label{app:3d_defo}

\begin{figure}[h!]
    \centering
    \includegraphics[width=1\linewidth]{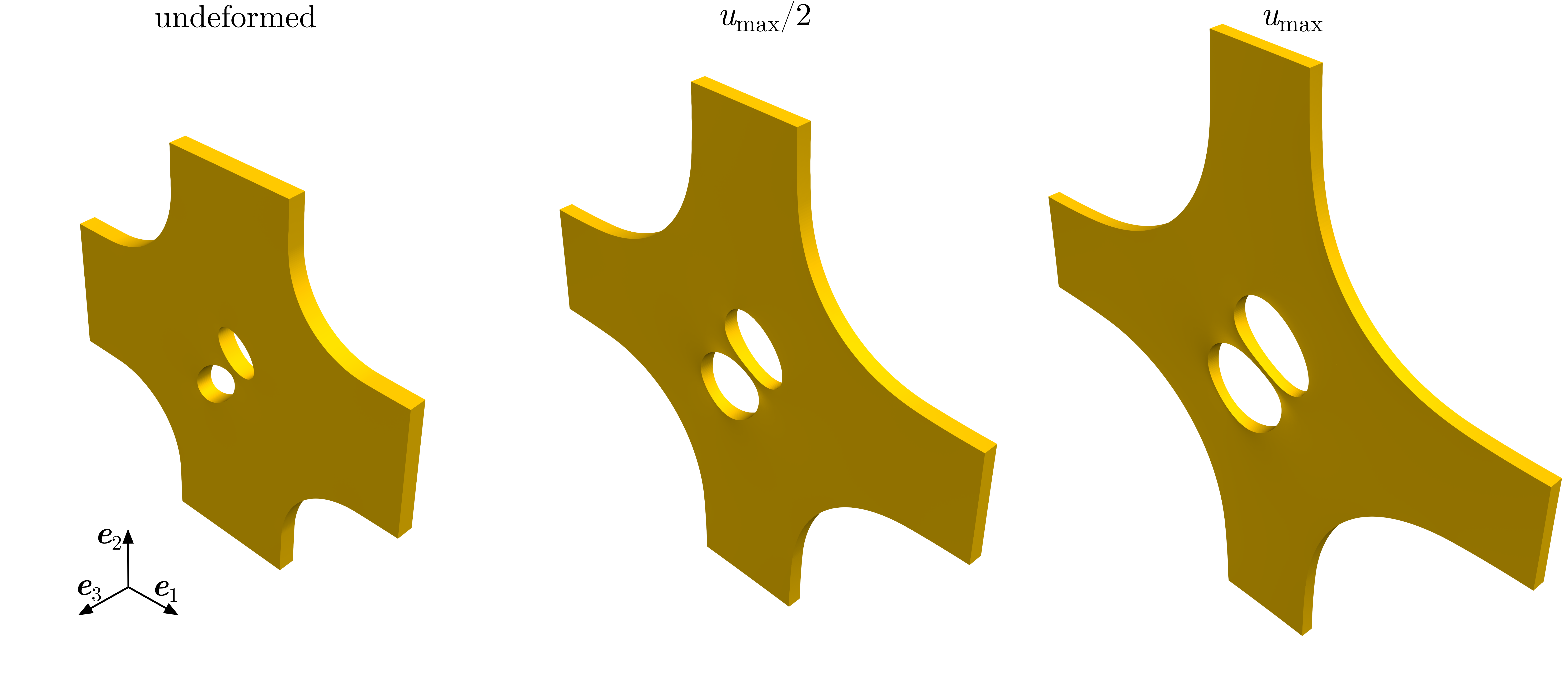}
    \caption{Visual of the three-dimensional simulation of the specimen with a thickness of $10\,\mathrm{mm}$ with maximum applied displacement $u_\mathrm{max} = 60\,\mathrm{mm}$ at each of the arms.}
    \label{fig:3D_defo_visual}
\end{figure}

\subsection{Artificial noise}
\label{app:noise}
Following~\cite{linden2025}, noise is superimposed on the displacement field by means of normalized Gaussian random fields $g_a$, generated on a regular rectilinear $2048 \times 2048$ grid encompassing the specimen and evaluated at the mesh nodes through bilinear interpolation.
\begin{figure}[h]
    \centering
    \includegraphics[width=1\linewidth, trim={0 1cm 0 1.cm},clip]{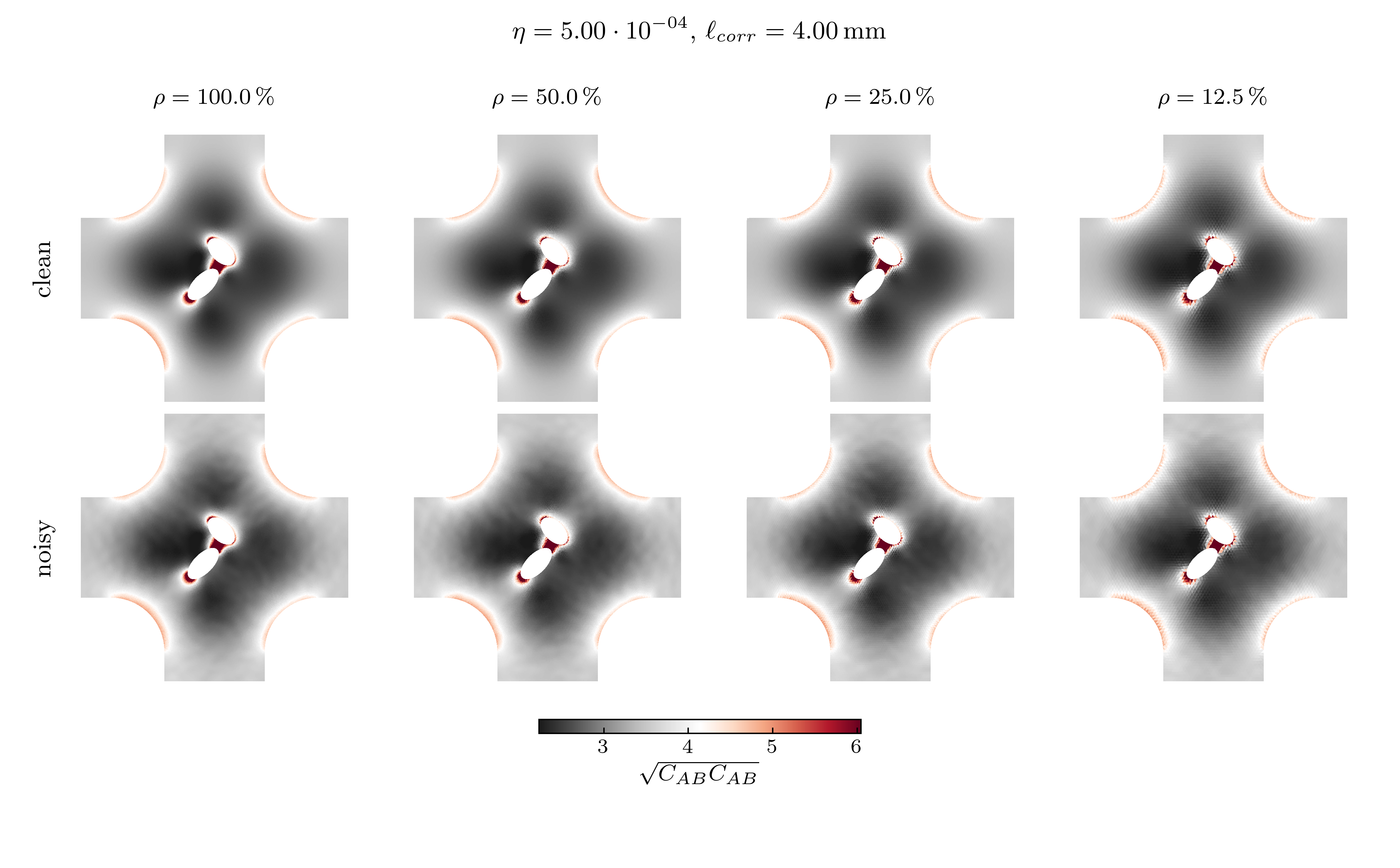}
    \caption{Effective noisy deformation field from adding noise to the displacement field.
            Norm of the in-plane components of the right Cauchy-Green deformation tensor at applied deformation boundary condition of $u_\mathrm{max}=60\,\mathrm{mm}$, depicted over the reference configuration.
            The noise applied to the displacement fields of the 3D simulations corresponds to $\rho=100\%$.}
    \label{fig:field_noise}
\end{figure}
With the fluctuation magnitude $\eta = 5 \cdot 10^{-4}$, correlation length $l = 4$ mm and length scale $\Delta x = 200\,\mathrm{mm}$ of the specimen under investigation, the components of the synthetic displacement field are perturbed individually at every increment according to
\begin{equation}
    \prescript{n\!}{}{\tilde{u}_a^\alpha}
    = \prescript{n\!}{}{u_a^\alpha} + g_a(\ve X^\alpha)\, \eta\, \Delta x \comma
    \label{eq:noise_u}
\end{equation}
see Fig.~\ref{fig:field_noise}.

The global reaction forces are perturbed as well, so as to emulate the limited accuracy of the load cells.
Since these are not field quantities, a simplified approach suffices: relative noise is applied by multiplication with samples from a uniform distribution $\mathcal{U}[1-\vartheta; 1+\vartheta]$, with $\vartheta = 10^{-4}$ throughout this work, i.e.,
\begin{equation}
    \prescript{n\!}{}{\tilde{F}_a^\varXi}
    = \prescript{n}{}{n}_{F^\varXi} \prescript{n\!}{}{F_a^\varXi} \quad 
    \mathrm{with} \quad
      \prescript{n}{}{n}_{F^\varXi} \sim \mathcal{U}[1-\vartheta; 1+\vartheta] \point
    \label{eq:noise_F}
\end{equation}
As the reaction forces of the cross-shaped specimen would be measured by two load cells in a real experiment, identical noise is applied to opposing boundaries, i.e., $\prescript{n}{}{n}_{F^1}  = \prescript{n}{}{n}_{F^3}  $ and $\prescript{n}{}{n}_{F^2}   = \prescript{n}{}{n}_{F^4}  $, for sequential numbering of the four boundaries.

\bibliographystyle{abbrv}  
{\footnotesize \bibliography{bib_vfmpann.bib}}	

@article{abbasi2026a,
  title = {Discovery of {{Hyperelastic Constitutive Laws}} from {{Experimental Data}} with {{EUCLID}}},
  author = {Abbasi, A. and Ricci, M. and Carrara, P. and Flaschel, M. and Kumar, S. and Marfia, S. and De Lorenzis, L.},
  year = 2026,
  month = jun,
  journal = {Experimental Mechanics},
  volume = {66},
  number = {5},
  pages = {877--908},
  issn = {0014-4851, 1741-2765},
  doi = {10.1007/s11340-026-01290-6},
  urldate = {2026-07-31},
  langid = {english},
}

@article{aldakheel2025,
  title = {Physics-Based Machine Learning for Computational Fracture Mechanics},
  author = {Aldakheel, Fadi and Elsayed, Elsayed S. and Heider, Yousef and Weeger, Oliver},
  year = 2025,
  month = jun,
  journal = {Machine Learning for Computational Science and Engineering},
  volume = {1},
  number = {1},
  pages = {18},
  issn = {3005-1428, 3005-1436},
  doi = {10.1007/s44379-025-00019-x},
  urldate = {2026-01-26},
  langid = {english},
}

@misc{alheit2026,
  title = {{{CANN-EUCLID}}: Unsupervised Constitutive Artificial Neural Network Model Discovery from Full-Field Data},
  shorttitle = {{{CANN-EUCLID}}},
  author = {Alheit, Benjamin and Kumar, Siddhant and Peirlinck, Mathias},
  year = 2026,
  month = jun,
  number = {arXiv:2606.14565},
  eprint = {2606.14565},
  primaryclass = {cs.CE},
  publisher = {arXiv},
  doi = {10.48550/arXiv.2606.14565},
  urldate = {2026-06-16},
  archiveprefix = {arXiv},
  langid = {english},
}

@inproceedings{amos,
  title = {Input {{Convex Neural Networks}}},
  author = {Amos, Brandon and Xu, Lei and Kolter, J Zico},
  year = {2017},
  booktitle = {Proceedings of the 34th {{International Conference}} on {{Machine Learning}}},
  volume = {70},
  series = {Proceedings of Machine Learning Research},
  pages = {146--155},
  publisher = {PMLR},
  langid = {english},
}

@article{arruda1993,
  title = {A Three-Dimensional Constitutive Model for the Large Stretch Behavior of Rubber Elastic Materials},
  author = {Arruda, Ellen M. and Boyce, Mary C.},
  year = 1993,
  month = feb,
  journal = {Journal of the Mechanics and Physics of Solids},
  volume = {41},
  number = {2},
  pages = {389--412},
  issn = {00225096},
  doi = {10.1016/0022-5096(93)90013-6},
  urldate = {2026-01-20},
  copyright = {https://www.elsevier.com/tdm/userlicense/1.0/},
  langid = {english},
}

@article{asad2022,
  title = {A Mechanics-Informed Artificial Neural Network Approach in Data-Driven Constitutive Modeling},
  author = {As'ad, Faisal and Avery, Philip and Farhat, Charbel},
  year = 2022,
  journal = {International Journal for Numerical Methods in Engineering},
  volume = {123},
  number = {12},
  pages = {2738--2759},
  issn = {1097-0207},
  doi = {10.1002/nme.6957},
  urldate = {2025-04-01},
  langid = {english},
}

@article{avril2008,
  title = {Overview of {{Identification Methods}} of {{Mechanical Parameters Based}} on {{Full-field Measurements}}},
  author = {Avril, St{\'e}phane and Bonnet, Marc and Bretelle, Anne-Sophie and Gr{\'e}diac, Michel and Hild, Fran{\c c}ois and Ienny, Patrick and Latourte, F{\'e}lix and Lemosse, Didier and Pagano, St{\'e}phane and Pagnacco, Emmanuel and Pierron, Fabrice},
  year = 2008,
  month = aug,
  journal = {Experimental Mechanics},
  volume = {48},
  number = {4},
  pages = {381--402},
  issn = {0014-4851, 1741-2765},
  doi = {10.1007/s11340-008-9148-y},
  urldate = {2026-03-06},
  copyright = {http://www.springer.com/tdm},
  langid = {english},
}

@article{bahmani2024,
  title = {Physics-constrained Symbolic Model Discovery for Polyconvex Incompressible Hyperelastic Materials},
  author = {Bahmani, Bahador and Sun, WaiChing},
  year = 2024,
  month = aug,
  journal = {International Journal for Numerical Methods in Engineering},
  volume = {125},
  number = {15},
  pages = {e7473},
  issn = {0029-5981, 1097-0207},
  doi = {10.1002/nme.7473},
  urldate = {2026-01-26},
  langid = {english},
}

@article{ball1976,
  title = {Convexity Conditions and Existence Theorems in Nonlinear Elasticity},
  author = {Ball, John M.},
  year = 1976,
  month = dec,
  journal = {Archive for Rational Mechanics and Analysis},
  volume = {63},
  number = {4},
  pages = {337--403},
  issn = {0003-9527, 1432-0673},
  doi = {10.1007/BF00279992},
  urldate = {2024-09-11},
  copyright = {http://www.springer.com/tdm},
  langid = {english},
}

@article{belloni2019,
  title = {{{py2DIC}}: {{A New Free}} and {{Open Source Software}} for {{Displacement}} and {{Strain Measurements}} in the {{Field}} of {{Experimental Mechanics}}},
  shorttitle = {{{py2DIC}}},
  author = {Belloni, Valeria and Ravanelli, Roberta and Nascetti, Andrea and Di Rita, Martina and Mattei, Domitilla and Crespi, Mattia},
  year = 2019,
  month = sep,
  journal = {Sensors},
  volume = {19},
  number = {18},
  pages = {3832},
  issn = {1424-8220},
  doi = {10.3390/s19183832},
  urldate = {2025-11-04},
  langid = {english},
}

@article{benady2024,
  title = {Unsupervised Learning of History-Dependent Constitutive Material Laws with Thermodynamically-Consistent Neural Networks in the Modified {{Constitutive Relation Error}} Framework.},
  author = {Benady, Antoine and Baranger, Emmanuel and Chamoin, Ludovic},
  year = 2024,
  month = may,
  journal = {Computer Methods in Applied Mechanics and Engineering},
  volume = {425},
  pages = {116967},
  publisher = {Elsevier},
  doi = {10.1016/j.cma.2024.116967},
  urldate = {2025-04-16},
}

@article{benady2024a,
  title = {{{NN}}-{{mCRE}}: {{A}} Modified Constitutive Relation Error Framework for Unsupervised Learning of Nonlinear State Laws with Physics-augmented Neural Networks},
  shorttitle = {{{NN}}-{{mCRE}}},
  author = {Benady, Antoine and Baranger, Emmanuel and Chamoin, Ludovic},
  year = 2024,
  month = apr,
  journal = {International Journal for Numerical Methods in Engineering},
  volume = {125},
  number = {8},
  pages = {e7439},
  issn = {0029-5981, 1097-0207},
  doi = {10.1002/nme.7439},
  urldate = {2026-07-14},
  langid = {english},
}

@book{bertram1989,
  title = {{Axiomatische Einf\"uhrung in die Kontinuumsmechanik}},
  author = {Bertram, Albrecht},
  year = 1989,
  publisher = {BI-Wiss.-Verl},
  address = {Mannheim},
  isbn = {978-3-411-14031-2},
  langid = {german},
}

@article{bock2019,
  title = {A {{Review}} of the {{Application}} of {{Machine Learning}} and {{Data Mining Approaches}} in {{Continuum Materials Mechanics}}},
  author = {Bock, Frederic E. and Aydin, Roland C. and Cyron, Christian J. and Huber, Norbert and Kalidindi, Surya R. and Klusemann, Benjamin},
  year = 2019,
  month = may,
  journal = {Frontiers in Materials},
  volume = {6},
  pages = {110},
  issn = {2296-8016},
  doi = {10.3389/fmats.2019.00110},
  urldate = {2026-02-25},
  langid = {english},
}

@article{bourdyot2026,
  title = {Learning a Hyperelastic Constitutive Model from {{3D}} Experimental Data},
  author = {Bourdyot, M. and Compans, M. and Langlois, R. and Smaniotto, B. and Baranger, E. and Jailin, C.},
  year = 2026,
  month = mar,
  journal = {Computer Methods in Applied Mechanics and Engineering},
  volume = {450},
  pages = {118592},
  issn = {00457825},
  doi = {10.1016/j.cma.2025.118592},
  urldate = {2026-07-14},
  langid = {english},
}

@article{chaurasiya2026,
  title = {Hetero-{{EUCLID}}: {{Interpretable}} Model Discovery for Heterogeneous Hyperelastic Materials Using Stress-Unsupervised Learning},
  shorttitle = {Hetero-{{EUCLID}}},
  author = {Chaurasiya, Kanhaiya Lal and Dutta, Saurav and Kumar, Siddhant and Joshi, Akshay},
  year = 2026,
  month = apr,
  journal = {Computer Methods in Applied Mechanics and Engineering},
  volume = {452},
  pages = {118729},
  issn = {00457825},
  doi = {10.1016/j.cma.2026.118729},
  urldate = {2026-07-14},
  langid = {english},
}

@article{chen2022,
  title = {Polyconvex Neural Networks for Hyperelastic Constitutive Models: {{A}} Rectification Approach},
  shorttitle = {Polyconvex Neural Networks for Hyperelastic Constitutive Models},
  author = {Chen, Peiyi and Guilleminot, Johann},
  year = 2022,
  month = oct,
  journal = {Mechanics Research Communications},
  volume = {125},
  pages = {103993},
  issn = {0093-6413},
  doi = {10.1016/j.mechrescom.2022.103993},
  urldate = {2026-09-02},
}

@article{chevalier2001,
  title = {Digital Image Correlation Used to Analyze the Multiaxial Behavior of Rubber-like Materials},
  author = {Chevalier, Luc and Calloch, Sylvain and Hild, Fran{\c c}ois and Marco, Yann},
  year = 2001,
  month = mar,
  journal = {European Journal of Mechanics - A/Solids},
  volume = {20},
  number = {2},
  pages = {169--187},
  issn = {09977538},
  doi = {10.1016/S0997-7538(00)01135-9},
  urldate = {2024-12-20},
  copyright = {https://www.elsevier.com/tdm/userlicense/1.0/},
  langid = {english},
}

@article{chrysochoos2010,
  title = {Use of {{Full-Field Digital Image Correlation}} and {{Infrared Thermography Measurements}} for the {{Thermomechanical Analysis}} of {{Material Behaviour}}},
  author = {Chrysochoos, A. and Huon, V. and Jourdan, F. and Muracciole, J.-M. and Peyroux, R. and Wattrisse, B.},
  year = 2010,
  month = feb,
  journal = {Strain},
  volume = {46},
  number = {1},
  pages = {117--130},
  issn = {00392103, 14751305},
  doi = {10.1111/j.1475-1305.2009.00635.x},
  urldate = {2026-05-03},
  copyright = {http://doi.wiley.com/10.1002/tdm\_license\_1.1},
  langid = {english},
}

@book{ciarlet1988a,
  title = {Mathematical Elasticity},
  author = {Ciarlet, Philippe Gaston},
  year = 1988,
  series = {Studies in Mathematics and Its Applications},
  number = {v. 20, 27, 29},
  publisher = {{North-Holland Sole distributors for the U.S.A. and Canada, Elsevier Science Pub. Co}},
  address = {Amsterdam New York New York, N.Y., U.S.A},
  isbn = {978-0-444-70259-3 978-0-444-82570-4 978-0-444-82891-0},
  langid = {english},
  lccn = {531.381},
}

@article{claire2004,
  title = {A Finite Element Formulation to Identify Damage Fields: The Equilibrium Gap Method},
  shorttitle = {A Finite Element Formulation to Identify Damage Fields},
  author = {Claire, D. and Hild, F. and Roux, S.},
  year = 2004,
  month = sep,
  journal = {International Journal for Numerical Methods in Engineering},
  volume = {61},
  number = {2},
  pages = {189--208},
  issn = {0029-5981, 1097-0207},
  doi = {10.1002/nme.1057},
  urldate = {2026-05-21},
  copyright = {http://onlinelibrary.wiley.com/termsAndConditions\#vor},
  langid = {english},
}

@article{dammass2025a,
  title = {Neural Networks Meet Phase-Field: {{A}} Hybrid Fracture Model},
  author = {Damma{\ss}, Franz and Kalina, Karl A. and K{\"a}stner, Markus},
  year = 2025,
  journal = {Computer Methods in Applied Mechanics and Engineering},
  volume = {440},
  pages = {117937},
  doi = {10.1016/j.cma.2025.117937},
  langid = {english},
}

@article{dammass2025b,
  title = {When Invariants Matter: {{The}} Role of {{I1}} and {{I2}} in Neural Network Models of Incompressible Hyperelasticity},
  shorttitle = {When Invariants Matter},
  author = {Damma{\ss}, Franz and Kalina, Karl A. and K{\"a}stner, Markus},
  year = 2025,
  month = nov,
  journal = {Mechanics of Materials},
  volume = {210},
  pages = {105443},
  issn = {01676636},
  doi = {10.1016/j.mechmat.2025.105443},
  urldate = {2025-10-01},
  langid = {english},
}

@article{dornheim2024,
  title = {Neural {{Networks}} for {{Constitutive Modeling}}: {{From Universal Function Approximators}} to {{Advanced Models}} and the {{Integration}} of {{Physics}}},
  shorttitle = {Neural {{Networks}} for {{Constitutive Modeling}}},
  author = {Dornheim, Johannes and Morand, Lukas and Nallani, Hemanth Janarthanam and Helm, Dirk},
  year = 2024,
  month = mar,
  journal = {Archives of Computational Methods in Engineering},
  volume = {31},
  number = {2},
  pages = {1097--1127},
  issn = {1134-3060, 1886-1784},
  doi = {10.1007/s11831-023-10009-y},
  urldate = {2026-01-15},
  langid = {english},
}

@article{ferreira2026,
  title = {Automatically {{Differentiable Model Updating}} ({{ADiMU}}): {{Conventional}}, Hybrid, and Neural Network Material Model Discovery Including History-Dependency},
  shorttitle = {Automatically {{Differentiable Model Updating}} ({{ADiMU}})},
  author = {Ferreira, Bernardo P. and Bessa, Miguel A.},
  year = 2026,
  month = jan,
  journal = {Journal of the Mechanics and Physics of Solids},
  volume = {206},
  pages = {106408},
  issn = {00225096},
  doi = {10.1016/j.jmps.2025.106408},
  urldate = {2026-07-14},
  langid = {english},
}

@article{flaschel2021,
  title = {Unsupervised Discovery of Interpretable Hyperelastic Constitutive Laws},
  author = {Flaschel, Moritz and Kumar, Siddhant and De Lorenzis, Laura},
  year = 2021,
  month = aug,
  journal = {Computer Methods in Applied Mechanics and Engineering},
  volume = {381},
  pages = {113852},
  issn = {00457825},
  doi = {10.1016/j.cma.2021.113852},
  urldate = {2025-03-31},
  langid = {english},
}

@article{flaschel2023,
  title = {Automated Discovery of Generalized Standard Material Models with {{EUCLID}}},
  author = {Flaschel, Moritz and Kumar, Siddhant and De Lorenzis, Laura},
  year = 2023,
  month = feb,
  journal = {Computer Methods in Applied Mechanics and Engineering},
  volume = {405},
  pages = {115867},
  issn = {00457825},
  doi = {10.1016/j.cma.2022.115867},
  urldate = {2024-06-30},
  langid = {english},
}

@article{flory1961,
  title = {Thermodynamic Relations for High Elastic Materials},
  author = {Flory, P. J.},
  year = 1961,
  journal = {Transactions of the Faraday Society},
  volume = {57},
  pages = {829--838},
  issn = {0014-7672},
  doi = {10.1039/tf9615700829},
  urldate = {2026-07-16},
  langid = {english},
}

@article{friedrichs2026,
  title = {Precise, Efficient and Flexible Modeling of Crystallizing Elastomers Based on Physics-Augmented Neural Networks},
  author = {Friedrichs, Konrad and Damma{\ss}, Franz and Kalina, Karl A. and K{\"a}stner, Markus},
  year = 2026,
  month = jun,
  journal = {Computer Methods in Applied Mechanics and Engineering},
  volume = {455},
  pages = {118852},
  issn = {00457825},
  doi = {10.1016/j.cma.2026.118852},
  urldate = {2026-03-13},
  langid = {english},
}

@article{fuhg2022,
  title = {Learning Hyperelastic Anisotropy from Data via a Tensor Basis Neural Network},
  author = {Fuhg, J.N. and Bouklas, N. and Jones, R.E.},
  year = 2022,
  month = nov,
  journal = {Journal of the Mechanics and Physics of Solids},
  volume = {168},
  pages = {105022},
  issn = {00225096},
  doi = {10.1016/j.jmps.2022.105022},
  urldate = {2026-09-02},
  langid = {english},
}

@article{fuhg2023,
  title = {Modular Machine Learning-Based Elastoplasticity: {{Generalization}} in the Context of Limited Data},
  shorttitle = {Modular Machine Learning-Based Elastoplasticity},
  author = {Fuhg, Jan Niklas and Hamel, Craig M. and Johnson, Kyle and Jones, Reese and Bouklas, Nikolaos},
  year = 2023,
  month = mar,
  journal = {Computer Methods in Applied Mechanics and Engineering},
  volume = {407},
  pages = {115930},
  issn = {00457825},
  doi = {10.1016/j.cma.2023.115930},
  urldate = {2026-09-02},
  langid = {english},
}

@article{fuhg2025,
  title = {A {{Review}} on {{Data-Driven Constitutive Laws}} for {{Solids}}},
  author = {Fuhg, Jan N. and Anantha Padmanabha, Govinda and Bouklas, Nikolaos and Bahmani, Bahador and Sun, WaiChing and Vlassis, Nikolaos N. and Flaschel, Moritz and Carrara, Pietro and De Lorenzis, Laura},
  year = 2025,
  month = apr,
  journal = {Archives of Computational Methods in Engineering},
  volume = {32},
  number = {3},
  pages = {1841--1883},
  issn = {1134-3060, 1886-1784},
  doi = {10.1007/s11831-024-10196-2},
  urldate = {2026-01-15},
  langid = {english},
}

@article{gavris2025,
  title = {Discovering Neural Elastoplasticity from Kinematic Observations},
  author = {Gavris, Georgios Barkoulis and Sun, WaiChing},
  year = 2025,
  month = sep,
  journal = {Proceedings of the National Academy of Sciences},
  volume = {122},
  number = {38},
  pages = {e2508732122},
  issn = {0027-8424, 1091-6490},
  doi = {10.1073/pnas.2508732122},
  urldate = {2026-07-14},
  langid = {english},
}

@article{gavris2026,
  title = {Discovering Neural Cohesive Zone Laws from Displacement Fields},
  author = {Gavris, Georgios Barkoulis and Sun, WaiChing},
  year = 2026,
  month = apr,
  journal = {Computer Methods in Applied Mechanics and Engineering},
  volume = {452},
  pages = {118733},
  issn = {00457825},
  doi = {10.1016/j.cma.2026.118733},
  urldate = {2026-07-14},
  langid = {english},
}

@article{geuken2025a,
  title = {A Novel Neural Network for Isotropic Polyconvex Hyperelasticity Satisfying the Universal Approximation Theorem},
  author = {Geuken, Gian-Luca and Kurzeja, Patrick and Wiedemann, David and Mosler, J{\"o}rn},
  year = 2025,
  month = oct,
  journal = {Journal of the Mechanics and Physics of Solids},
  volume = {203},
  pages = {106209},
  issn = {00225096},
  doi = {10.1016/j.jmps.2025.106209},
  urldate = {2026-08-28},
  langid = {english},
}

@article{ghaboussi1991,
  title = {Knowledge-{{Based Modeling}} of {{Material Behavior}} with {{Neural Networks}}},
  author = {Ghaboussi, J. and Garrett, J. H. and Wu, X.},
  year = 1991,
  month = jan,
  journal = {Journal of Engineering Mechanics},
  volume = {117},
  number = {1},
  pages = {132--153},
  issn = {0733-9399, 1943-7889},
  doi = {10.1061/(ASCE)0733-9399(1991)117:1(132)},
  urldate = {2025-08-31},
  langid = {english},
}

@article{grediac1989,
  title = {Principe {{Des Travaux Virtuels}} et {{Identification}}},
  author = {Gr{\'e}diac, M},
  year = 1989,
  journal = {Comptes rendus de l'Acad\'emie des sciences. S\'erie 2, M\'ecanique, Physique, Chimie, Sciences de l'univers, Sciences de la Terre},
  volume = {309},
  number = {1},
  pages = {1--5}
}

@book{haupt2002,
  title = {Continuum {{Mechanics}} and {{Theory}} of {{Materials}}},
  author = {Haupt, Peter},
  year = 2002,
  series = {Advanced {{Texts}} in {{Physics}}},
  publisher = {Springer Berlin Heidelberg},
  address = {Berlin, Heidelberg},
  doi = {10.1007/978-3-662-04775-0},
  urldate = {2025-05-16},
  copyright = {http://www.springer.com/tdm},
  isbn = {978-3-642-07718-0 978-3-662-04775-0},
  langid = {english},
}

@article{jadoon2025a,
  title = {Inverse Design of Anisotropic Microstructures Using Physics-Augmented Neural Networks},
  author = {Jadoon, Asghar A. and Kalina, Karl A. and Rausch, Manuel K. and Jones, Reese and Fuhg, Jan Niklas},
  year = 2025,
  month = oct,
  journal = {Journal of the Mechanics and Physics of Solids},
  volume = {203},
  pages = {106161},
  issn = {0022-5096},
  doi = {10.1016/j.jmps.2025.106161},
  urldate = {2026-09-02},
}

@article{jones1975,
  title = {The Properties of Rubber in Pure Homogeneous Strain},
  author = {Jones, D F and Treloar, L R G},
  year = 1975,
  month = aug,
  journal = {Journal of Physics D: Applied Physics},
  volume = {8},
  number = {11},
  pages = {1285--1304},
  issn = {0022-3727, 1361-6463},
  doi = {10.1088/0022-3727/8/11/007},
  urldate = {2024-12-09},
  langid = {english},
}

@article{kalina2023,
  title = {{{FEANN}}: An Efficient Data-Driven Multiscale Approach Based on Physics-Constrained Neural Networks and Automated Data Mining},
  shorttitle = {{{FE}}\$\$\textbraceleft\textbraceright\textasciicircum\textbackslash textrm\textbraceleft{{ANN}}\textbraceright\$\$},
  author = {Kalina, Karl A. and Linden, Lennart and Brummund, J{\"o}rg and K{\"a}stner, Markus},
  year = 2023,
  month = may,
  journal = {Computational Mechanics},
  volume = {71},
  number = {5},
  pages = {827--851},
  issn = {0178-7675, 1432-0924},
  doi = {10.1007/s00466-022-02260-0},
  urldate = {2025-04-16},
  langid = {english},
}

@article{kalina2024,
  title = {Neural Network-Based Multiscale Modeling of Finite Strain Magneto-Elasticity with Relaxed Convexity Criteria},
  author = {Kalina, Karl A. and Gebhart, Philipp and Brummund, J{\"o}rg and Linden, Lennart and Sun, WaiChing and K{\"a}stner, Markus},
  year = 2024,
  month = mar,
  journal = {Computer Methods in Applied Mechanics and Engineering},
  volume = {421},
  pages = {116739},
  issn = {00457825},
  doi = {10.1016/j.cma.2023.116739},
  urldate = {2024-06-30},
  langid = {english},
}

@article{kawabata1981,
  title = {Experimental Survey of the Strain Energy Density Function of Isoprene Rubber Vulcanizate},
  author = {Kawabata, S. and Matsuda, M. and Tei, K. and Kawai, H.},
  year = 1981,
  month = jan,
  journal = {Macromolecules},
  volume = {14},
  number = {1},
  pages = {154--162},
  issn = {0024-9297, 1520-5835},
  doi = {10.1021/ma50002a032},
  urldate = {2025-11-04},
  langid = {english},
}

@article{klein2022,
  title = {Polyconvex Anisotropic Hyperelasticity with Neural Networks},
  author = {Klein, Dominik K. and Fern{\'a}ndez, Mauricio and Martin, Robert J. and Neff, Patrizio and Weeger, Oliver},
  year = 2022,
  month = feb,
  journal = {Journal of the Mechanics and Physics of Solids},
  volume = {159},
  eprint = {2106.14623},
  primaryclass = {cond-mat},
  pages = {104703},
  issn = {00225096},
  doi = {10.1016/j.jmps.2021.104703},
  urldate = {2025-04-16},
  archiveprefix = {arXiv},
}

@misc{knipper2026,
  title = {Finite {{Element-Based Material Learning}} via {{Automatic Differentiation}}: {{Learning}} Constitutive Neural Network Models from Full-Field Deformation Data},
  shorttitle = {Finite {{Element-Based Material Learning}} via {{Automatic Differentiation}}},
  author = {Knipper, Matthias and Ji, Chenyi and Brand, Malte and Linka, Kevin},
  year = 2026,
  month = may,
  number = {arXiv:2606.05199},
  eprint = {2606.05199},
  primaryclass = {physics.comp-ph},
  publisher = {arXiv},
  doi = {10.48550/arXiv.2606.05199},
  urldate = {2026-07-14},
  archiveprefix = {arXiv},
  langid = {english},
}

@article{li2026,
  title = {Learning Anisotropic Hyperelasticity with an Unsupervised Symmetry-Aware Equilibrium-Based Neural Network},
  author = {Li, Shun and Li, Lingfeng and Chen, Chang Qing},
  year = 2026,
  month = jun,
  journal = {Computer Methods in Applied Mechanics and Engineering},
  volume = {455},
  pages = {118911},
  issn = {00457825},
  doi = {10.1016/j.cma.2026.118911},
  urldate = {2026-07-14},
  langid = {english},
}

@article{linden2023,
  title = {Neural Networks Meet Hyperelasticity: {{A}} Guide to Enforcing Physics},
  shorttitle = {Neural Networks Meet Hyperelasticity},
  author = {Linden, Lennart and Klein, Dominik K. and Kalina, Karl A. and Brummund, J{\"o}rg and Weeger, Oliver and K{\"a}stner, Markus},
  year = 2023,
  month = oct,
  journal = {Journal of the Mechanics and Physics of Solids},
  volume = {179},
  pages = {105363},
  issn = {00225096},
  doi = {10.1016/j.jmps.2023.105363},
  urldate = {2024-06-30},
  langid = {english},
}

@article{linden2025,
  title = {A Dual-Stage Constitutive Modeling Framework Based on Finite Strain Data-Driven Identification and Physics-Augmented Neural Networks},
  author = {Linden, Lennart and Kalina, Karl A. and Brummund, J{\"o}rg and Riemer, Brain and K{\"a}stner, Markus},
  year = 2025,
  month = dec,
  journal = {Computer Methods in Applied Mechanics and Engineering},
  volume = {447},
  pages = {118289},
  issn = {00457825},
  doi = {10.1016/j.cma.2025.118289},
  urldate = {2026-06-24},
  langid = {english},
}

@article{linka2021,
  title = {Constitutive Artificial Neural Networks: {{A}} Fast and General Approach to Predictive Data-Driven Constitutive Modeling by Deep Learning},
  shorttitle = {Constitutive Artificial Neural Networks},
  author = {Linka, Kevin and Hillg{\"a}rtner, Markus and Abdolazizi, Kian P. and Aydin, Roland C. and Itskov, Mikhail and Cyron, Christian J.},
  year = 2021,
  month = mar,
  journal = {Journal of Computational Physics},
  volume = {429},
  pages = {110010},
  issn = {0021-9991},
  doi = {10.1016/j.jcp.2020.110010},
  urldate = {2025-04-16},
}

@article{linka2023,
  title = {A New Family of {{Constitutive Artificial Neural Networks}} towards Automated Model Discovery},
  author = {Linka, Kevin and Kuhl, Ellen},
  year = 2023,
  month = jan,
  journal = {Computer Methods in Applied Mechanics and Engineering},
  volume = {403},
  pages = {115731},
  issn = {00457825},
  doi = {10.1016/j.cma.2022.115731},
  urldate = {2026-07-14},
  langid = {english},
}

@article{lourenco2024,
  title = {An Indirect Training Approach for Implicit Constitutive Modelling Using Recurrent Neural Networks and the Virtual Fields Method},
  author = {Louren{\c c}o, R{\'u}ben and Georgieva, Petia and Cueto, Elias and {Andrade-Campos}, A.},
  year = 2024,
  month = may,
  journal = {Computer Methods in Applied Mechanics and Engineering},
  volume = {425},
  pages = {116961},
  issn = {00457825},
  doi = {10.1016/j.cma.2024.116961},
  urldate = {2026-07-14},
  langid = {english},
}

@article{marckmann2006,
  title = {Comparison of {{Hyperelastic Models}} for {{Rubber-Like Materials}}},
  author = {Marckmann, G. and Verron, E.},
  year = 2006,
  month = nov,
  journal = {Rubber Chemistry and Technology},
  volume = {79},
  number = {5},
  pages = {835--858},
  issn = {1943-4804, 0035-9475},
  doi = {10.5254/1.3547969},
  urldate = {2025-04-07},
  langid = {english},
}

@book{marsden1984,
  title = {Mathematical {{Foundations}} of {{Elasticity}}},
  author = {Marsden, Jerrold E. and Hughes, Thomas J. R.},
  year = 1983,
  publisher = {Prentice-Hall},
  address = {Englewood Cliffs, {NJ}},
  series = {Prentice-{Hall} Civil Engineering and Engineering Mechanics Series},
  isbn = {978-0-13-561076-3},
  langid = {english},
}

@article{masi2021,
  title = {Thermodynamics-Based {{Artificial Neural Networks}} for Constitutive Modeling},
  author = {Masi, Filippo and Stefanou, Ioannis and Vannucci, Paolo and {Maffi-Berthier}, Victor},
  year = 2021,
  month = feb,
  journal = {Journal of the Mechanics and Physics of Solids},
  volume = {147},
  pages = {104277},
  issn = {00225096},
  doi = {10.1016/j.jmps.2020.104277},
  urldate = {2026-09-02},
  langid = {english},
}

@article{meng2025,
  title = {Machine-Learning-Based Virtual Fields Method: {{Application}} to Anisotropic Hyperelasticity},
  shorttitle = {Machine-Learning-Based Virtual Fields Method},
  author = {Meng, Shuangshuang and Yousefi, Ali Akbar Karkhaneh and Avril, St{\'e}phane},
  year = 2025,
  month = feb,
  journal = {Computer Methods in Applied Mechanics and Engineering},
  volume = {434},
  pages = {117580},
  issn = {00457825},
  doi = {10.1016/j.cma.2024.117580},
  urldate = {2026-07-14},
  langid = {english},
}

@misc{moon2026,
  title = {Physics-{{Informed Discovery}} of {{Yield Functions}} in {{Plasticity}} via {{Convex Neural Representations}}},
  author = {Moon, Hyeonbin and Cho, Donghyuk and Yu, Jecheon and Yoon, Jeong Whan and Ryu, Seunghwa},
  year = 2026,
  month = jun,
  number = {arXiv:2606.19375},
  eprint = {2606.19375},
  primaryclass = {cs.LG},
  publisher = {arXiv},
  doi = {10.48550/arXiv.2606.19375},
  urldate = {2026-07-14},
  archiveprefix = {arXiv},
}

@article{pascon2019,
  title = {Large Deformation Analysis of Plane-Stress Hyperelastic Problems via Triangular Membrane Finite Elements},
  author = {Pascon, Jo{\~a}o Paulo},
  year = 2019,
  month = sep,
  journal = {International Journal of Advanced Structural Engineering},
  volume = {11},
  number = {3},
  pages = {331--350},
  issn = {2008-3556, 2008-6695},
  doi = {10.1007/s40091-019-00234-w},
  urldate = {2025-09-29},
  langid = {english},
}

@article{peirlinck2024,
  title = {On Automated Model Discovery and a Universal Material Subroutine for Hyperelastic Materials},
  author = {Peirlinck, Mathias and Linka, Kevin and Hurtado, Juan A. and Kuhl, Ellen},
  year = 2024,
  month = jan,
  journal = {Computer Methods in Applied Mechanics and Engineering},
  volume = {418},
  pages = {116534},
  issn = {0045-7825},
  doi = {10.1016/j.cma.2023.116534},
  urldate = {2026-09-02},
}

@book{pierron2012,
  title = {The {{Virtual Fields Method}}},
  author = {Pierron, Fabrice and Grediac, Michel},
  year = 2012,
  publisher = {Springer New York},
  doi = {10.1007/978-1-4614-1824-5},
  isbn = {978-1-4614-1823-8 978-1-4614-1824-5}
}

@article{pierron2021,
  title = {Towards {{Material Testing}} 2.0. {{A}} Review of Test Design for Identification of Constitutive Parameters from Full-field Measurements},
  author = {Pierron, F. and Gr{\'e}diac, M.},
  year = 2021,
  month = feb,
  journal = {Strain},
  volume = {57},
  number = {1},
  pages = {e12370},
  issn = {0039-2103, 1475-1305},
  doi = {10.1111/str.12370},
  urldate = {2026-07-02},
  langid = {english},
}

@article{raissi2019,
  title = {Physics-Informed Neural Networks: {{A}} Deep Learning Framework for Solving Forward and Inverse Problems Involving Nonlinear Partial Differential Equations},
  shorttitle = {Physics-Informed Neural Networks},
  author = {Raissi, M. and Perdikaris, P. and Karniadakis, G.E.},
  year = 2019,
  month = feb,
  journal = {Journal of Computational Physics},
  volume = {378},
  pages = {686--707},
  issn = {00219991},
  doi = {10.1016/j.jcp.2018.10.045},
  urldate = {2026-09-02},
  langid = {english},
}

@article{ricker2023,
  title = {Systematic {{Fitting}} and {{Comparison}} of {{Hyperelastic Continuum Models}} for {{Elastomers}}},
  author = {Ricker, Alexander and Wriggers, Peter},
  year = 2023,
  month = apr,
  journal = {Archives of Computational Methods in Engineering},
  volume = {30},
  number = {3},
  pages = {2257--2288},
  issn = {1134-3060, 1886-1784},
  doi = {10.1007/s11831-022-09865-x},
  urldate = {2024-12-09},
  langid = {english},
}

@article{riemer2026,
  title = {Construction of Minimal Integrity Bases for Anisotropic Hyperelasticity via Structural Tensors},
  author = {Riemer, Brain M. and Brummund, J{\"o}rg and Kalina, Karl A. and Milor, Abel H.G. and Damma{\ss}, Franz and K{\"a}stner, Markus},
  year = 2026,
  month = oct,
  journal = {Journal of the Mechanics and Physics of Solids},
  volume = {216},
  pages = {106763},
  issn = {00225096},
  doi = {10.1016/j.jmps.2026.106763},
  urldate = {2026-07-28},
  langid = {english},
}

@misc{riemer2026a,
  title = {Calibration of Neural Viscoelastic Models via Full-Field Data},
  author = {Riemer, Brain M. and K{\"a}stner, Markus and Kalina, Karl A.},
  year = 2026,
  month = sep,
  number = {arXiv:2609.03645},
  eprint = {2609.03645},
  primaryclass = {cs.CE},
  publisher = {arXiv},
  doi = {10.48550/arXiv.2609.03645},
  urldate = {2026-09-04},
  archiveprefix = {arXiv},
}

@article{romer2025,
  title = {Reduced and {{All-At-Once Approaches}} for {{Model Calibration}} and {{Discovery}} in {{Computational Solid Mechanics}}},
  author = {R{\"o}mer, Ulrich and Hartmann, Stefan and Tr{\"o}ger, Jendrik-Alexander and Anton, David and Wessels, Henning and Flaschel, Moritz and De Lorenzis, Laura},
  year = 2025,
  month = jul,
  journal = {Applied Mechanics Reviews},
  volume = {77},
  number = {4},
  pages = {040801},
  issn = {0003-6900, 2379-0407},
  doi = {10.1115/1.4066118},
  urldate = {2026-07-14},
  langid = {english},
}

@book{schroder2010,
  title = {Poly-, {{Quasi-}} and {{Rank-One Convexity}} in {{Applied Mechanics}}},
  editor = {Schr{\"o}der, J{\"o}rg and Neff, Patrizio and Maier, Giulio and Rammerstorfer, Franz G. and Salen{\c c}on, Jean and Schrefler, Bernhard and Serafini, Paolo},
  year = 2010,
  series = {{{CISM International Centre}} for {{Mechanical Sciences}}},
  volume = {516},
  publisher = {Springer Vienna},
  address = {Vienna},
  doi = {10.1007/978-3-7091-0174-2},
  urldate = {2025-09-01},
  copyright = {http://www.springer.com/tdm},
  isbn = {978-3-7091-0173-5 978-3-7091-0174-2},
  langid = {english},
}

@book{scott1992,
  title = {Multivariate Density Estimation: Theory, Practice, and Visualization},
  shorttitle = {Multivariate Density Estimation},
  author = {Scott, David W.},
  year = 1992,
  series = {Wiley Series in Probability and Mathematical Statistics},
  publisher = {J. Wiley},
  address = {New York Chichester Brisbane},
  isbn = {978-0-471-54770-9},
  langid = {english},
  lccn = {519.535}
}

@article{shi2025a,
  title = {Deep {{Learning}} without {{Stress Data}} on the {{Discovery}} of {{Multi-Regional Hyperelastic Properties}}},
  author = {Shi, Ruike and Yang, Haitian and Chen, Jianxu and Hackl, Klaus and Avril, St{\'e}phane and He, Yiqian},
  year = 2025,
  journal = {Computational Mechanics},
  volume = {76},
  pages = {117--146},
  doi = {10.1007/s00466-024-02591-0}
}

@book{silhavy1997,
  title = {The {{Mechanics}} and {{Thermodynamics}} of {{Continuous Media}}},
  author = {{\v S}ilhav{\'y}, Miroslav},
  year = 1997,
  publisher = {Springer Berlin Heidelberg},
  address = {Berlin, Heidelberg},
  doi = {10.1007/978-3-662-03389-0},
  urldate = {2025-03-11},
  copyright = {http://www.springer.com/tdm},
  isbn = {978-3-642-08204-7 978-3-662-03389-0},
  langid = {english},
}

@article{steinmann2012,
  title = {Hyperelastic Models for Rubber-like Materials: Consistent Tangent Operators and Suitability for {{Treloar}}'s Data},
  shorttitle = {Hyperelastic Models for Rubber-like Materials},
  author = {Steinmann, Paul and Hossain, Mokarram and Possart, Gunnar},
  year = 2012,
  month = sep,
  journal = {Archive of Applied Mechanics},
  volume = {82},
  number = {9},
  pages = {1183--1217},
  issn = {0939-1533, 1432-0681},
  doi = {10.1007/s00419-012-0610-z},
  urldate = {2024-06-27},
  copyright = {http://www.springer.com/tdm},
  langid = {english},
}

@article{tac2024,
  title = {Benchmarking Physics-Informed Frameworks for Data-Driven Hyperelasticity},
  author = {Ta{\c c}, Vahidullah and Linka, Kevin and {Sahli-Costabal}, Francisco and Kuhl, Ellen and Tepole, Adrian Buganza},
  year = 2024,
  month = jan,
  journal = {Computational Mechanics},
  volume = {73},
  number = {1},
  pages = {49--65},
  issn = {1432-0924},
  doi = {10.1007/s00466-023-02355-2},
  urldate = {2026-09-02},
  langid = {english},
}

@article{tac2026,
  title = {Fully Data-Driven Inverse Characterization of Heterogeneous Materials with Hyper-Network Neural {{ODEs}}},
  author = {Ta{\c c}, Vahidullah and {Amiri-Hezaveh}, Amirhossein and Bechtel, Grace N. and Loftin, Titus and Rausch, Manuel K. and Sahli Costabal, Francisco and Tepole, Adrian Buganza},
  year = 2026,
  month = mar,
  journal = {npj Computational Materials},
  volume = {12},
  number = {1},
  pages = {165},
  issn = {2057-3960},
  doi = {10.1038/s41524-026-02027-8},
  urldate = {2026-07-14},
  langid = {english},
}

@misc{tan2026,
  title = {Towards {{Rapid Constitutive Model Discovery}} from {{Multi-Modal Data}}: {{Physics Augmented Finite Element Model Updating}} ({{paFEMU}})},
  shorttitle = {Towards {{Rapid Constitutive Model Discovery}} from {{Multi-Modal Data}}},
  author = {Tan, Jingye and Padmanabha, Govinda Anantha and Yang, Steven J. and Bouklas, Nikolaos},
  year = 2026,
  publisher = {arXiv},
  doi = {10.48550/ARXIV.2604.07746},
  urldate = {2026-08-12},
  copyright = {arXiv.org perpetual, non-exclusive license},
  langid = {english},
}

@article{thakolkaran2022,
  title = {{{NN-EUCLID}}: {{Deep-learning}} Hyperelasticity without Stress Data},
  shorttitle = {{{NN-EUCLID}}},
  author = {Thakolkaran, Prakash and Joshi, Akshay and Zheng, Yiwen and Flaschel, Moritz and De Lorenzis, Laura and Kumar, Siddhant},
  year = 2022,
  month = dec,
  journal = {Journal of the Mechanics and Physics of Solids},
  volume = {169},
  pages = {105076},
  issn = {0022-5096},
  doi = {10.1016/j.jmps.2022.105076},
  urldate = {2025-03-31},
}

@article{thakolkaran2025,
  title = {Can {{KAN CANs}}? {{Input-convex Kolmogorov-Arnold Networks}} ({{KANs}}) as Hyperelastic Constitutive Artificial Neural Networks ({{CANs}})},
  shorttitle = {Can {{KAN CANs}}?},
  author = {Thakolkaran, Prakash and Guo, Yaqi and Saini, Shivam and Peirlinck, Mathias and Alheit, Benjamin and Kumar, Siddhant},
  year = 2025,
  month = aug,
  journal = {Computer Methods in Applied Mechanics and Engineering},
  volume = {443},
  pages = {118089},
  issn = {00457825},
  doi = {10.1016/j.cma.2025.118089},
  urldate = {2026-07-14},
  langid = {english},
}

@article{treloar1944,
  title = {Stress-Strain Data for Vulcanised Rubber under Various Types of Deformation},
  author = {Treloar, L. R. G.},
  year = 1944,
  journal = {Transactions of the Faraday Society},
  volume = {40},
  pages = {59--70},
  issn = {0014-7672},
  doi = {10.1039/tf9444000059},
  urldate = {2025-02-18},
  langid = {english},
}

@article{vijayakumaran2025,
  title = {Consistent Machine Learning for Topology Optimization with Microstructure-Dependent Neural Network Material Models},
  author = {Vijayakumaran, Harikrishnan and Russ, Jonathan B. and Paulino, Glaucio H. and Bessa, Miguel A.},
  year = 2025,
  month = mar,
  journal = {Journal of the Mechanics and Physics of Solids},
  volume = {196},
  pages = {106015},
  issn = {0022-5096},
  doi = {10.1016/j.jmps.2024.106015},
  urldate = {2026-09-02},
}

@article{wiesheier2024,
  title = {Versatile Data-Adaptive Hyperelastic Energy Functions for Soft Materials},
  author = {Wiesheier, Simon and {Moreno-Mateos}, Miguel Angel and Steinmann, Paul},
  year = 2024,
  month = oct,
  journal = {Computer Methods in Applied Mechanics and Engineering},
  volume = {430},
  pages = {117208},
  issn = {00457825},
  doi = {10.1016/j.cma.2024.117208},
  urldate = {2026-07-14},
  langid = {english},
}

@article{wiesheier2026,
  title = {Data-Adaptive Spline-Based Viscoelasticity for Soft Solids},
  author = {Wiesheier, Simon and {Moreno-Mateos}, Miguel Angel and Steinmann, Paul},
  year = 2026,
  month = apr,
  journal = {Computer Methods in Applied Mechanics and Engineering},
  volume = {451},
  pages = {118705},
  issn = {00457825},
  doi = {10.1016/j.cma.2025.118705},
  urldate = {2026-01-26},
  langid = {english},
}

@article{wu2025,
  title = {Learning the Physics-Consistent Material Behavior from Measurable Data via {{PDE-constrained}} Optimization},
  author = {Wu, Xinxin and Zhang, Yin and Mao, Sheng},
  year = 2025,
  month = mar,
  journal = {Computer Methods in Applied Mechanics and Engineering},
  volume = {437},
  pages = {117748},
  issn = {00457825},
  doi = {10.1016/j.cma.2025.117748},
  urldate = {2026-07-14},
  langid = {english},
}

@misc{baratta2023,
  title = {{{DOLFINx}}: {{The}} next Generation {{FEniCS}} Problem Solving Environment},
  shorttitle = {{{DOLFINx}}},
  author = {Baratta, Igor A. and Dean, Joseph P. and Dokken, J{\o}rgen S. and Habera, Michal and Hale, Jack S. and Richardson, Chris N. and Rognes, Marie E. and Scroggs, Matthew W. and Sime, Nathan and Wells, Garth N.},
  year = {2023},
  month = dec,
  publisher = {Zenodo},
  doi = {10.5281/zenodo.10447666},
  urldate = {2024-11-25},
  archiveprefix = {Zenodo},
  langid = {english},
}

@article{scroggs2022,
  title = {Basix: A Runtime Finite Element Basis Evaluation Library},
  shorttitle = {Basix},
  author = {Scroggs, Matthew W. and Baratta, Igor A. and Richardson, Chris N. and Wells, Garth N.},
  year = {2022},
  month = may,
  journal = {Journal of Open Source Software},
  volume = {7},
  number = {73},
  pages = {3982},
  issn = {2475-9066},
  doi = {10.21105/joss.03982},
  urldate = {2024-11-25},
  langid = {english},
}

@article{scroggs2022a,
  title = {Construction of {{Arbitrary Order Finite Element Degree-of-Freedom Maps}} on {{Polygonal}} and {{Polyhedral Cell Meshes}}},
  author = {Scroggs, Matthew W. and Dokken, J{\o}rgen S. and Richardson, Chris N. and Wells, Garth N.},
  year = {2022},
  month = may,
  journal = {ACM Trans. Math. Softw.},
  volume = {48},
  number = {2},
  pages = {18:1--18:23},
  issn = {0098-3500},
  doi = {10.1145/3524456},
  urldate = {2024-11-25},
}

@book{wriggers2008,
  title = {Nonlinear Finite Element Methods},
  author = {Wriggers, Peter},
  year = {2008},
  publisher = {Springer},
  address = {Berlin Heidelberg},
  isbn = {978-3-540-71000-4 978-3-642-09002-8},
  langid = {english},
}

\end{document}